\documentclass[11pt,a4paper]{article}
\usepackage[utf8]{inputenc}
\usepackage{amsmath}
\usepackage{amsfonts}
\usepackage{amssymb}
\usepackage{graphicx}
\usepackage{amsthm}
\usepackage{hyperref}
\usepackage[left=2.5cm,right=2.5cm,top=2cm,bottom=2cm]{geometry}
\usepackage{authblk}
\usepackage{framed}
\usepackage{listings}
\usepackage{color}

\usepackage{chngcntr}
\counterwithout{figure}{section}

\usepackage{setspace}

\usepackage{color}
\usepackage{amsfonts, amsmath, amsfonts, amssymb}
\usepackage{graphicx}
\usepackage{algorithm,algpseudocode}
\usepackage{esint}

\usepackage{caption}

\theoremstyle{note}% default

\theoremstyle{remark}
\newtheorem*{rem}{Remark}

\begin{document}

\title{Characterizing Complex Networks with Forman-Ricci curvature and associated geometric flows}
%% List of authors, with corresponding author marked by asterisk
\author{Melanie Weber$^{1,2}$, Emil Saucan$^{1,3}$and J{\"u}rgen Jost $^{1,4}$\\
%% Author affiliations
\small $^{1}$ Max-Planck-Institute for Mathematics in the Sciences; Inselstrasse 22, 04103 Leipzig, Germany \\
\small $^{2}$ Mathematical Institute, University of Leipzig; Augustusplatz 10, 04109 Leipzig, Germany \\
\small $^{3}$ Technion - Israel Institute of Technology; Haifa 32000, Israel\\
\small $^{4}$ Santa Fe Institute; 1399 Hyde Park Road Santa Fe, New Mexico 87501 USA\\
{\footnotesize (Contact: melweber@t-online.de, semil@ee.technion.ac.il, jost@mis.mpg.de)}
}
\maketitle

\begin{abstract}
{\noindent We introduce Forman-Ricci curvature and its corresponding flow as characteristics for complex networks attempting to extend the common approach of node-based network analysis by edge-based characteristics. Following a theoretical introduction and mathematical motivation, we apply the proposed network-analytic methods to static and dynamic complex networks and compare the results with established node-based characteristics. Our work suggests a number of applications for data mining, including denoising and clustering of experimental data, as well as extrapolation of network evolution.}
{Complex networks, Forman-Ricci-curvature, Ricci-flow, Laplacian flow, data mining}
%%%% If classification number provided then
\\
2000 Math Subject Classification: 05C82, 05C75, 05C21, 05C10

\end{abstract}

Contact: mw25@math.princeton.edu, semil@ee.technion.ac.il, jost@mis.mpg.de

\section{Introduction}

\noindent Network graphs have been perceived as practical models for complex systems for decades. With the rapid rise of data science since the late 1990s, they have become a widely used form of data representation that is easily storable and can be effectively analyzed with data mining methods. 

Complex networks represent a wide variety of data sets and systems, ranging from social interactions in large online social networks like Facebook and Twitter to neuronal activities in cortical graphs and genetic interactions in the human genome. Outside the social and natural sciences, networks are used as models for the spread of information in the world wide web and the shared use of energy ressources -- to name just a few examples. %\add[MW]{Do we need references here or is it okay to just name some examples?}

Since the early days of network sciences, there have been numerous attempts to characterize complex real-world networks with standard models that capture essential topological properties. Most notably are the models of P. Erdős and A. Rényi \cite{er1,er2}, D. Watts and S. Strogatz \cite{ws} and R. Albert and A. Barabási \cite{ab}. Their efforts were addressed with both praise (e.g. \cite{stanley,jeong2}) and concerns (e.g. \cite{arita_metabolic_2004, marcotte}) in the literature in terms of their capabilities as well as deficiencies in describing real world systems. 

Common among most network models is the focus on node degree distributions, clustering coefficiants and the average path length as the defining topological and geometrical properties \cite{stroegatz1,newman1}. More recently \cite{jost1, jost2}, the spectrum of the normalized graph Laplacian has been introduced as a useful classification scheme. In this work, we suggest additional edge-based characteristics based on combinatorial approaches, namely the Forman-Ricci curvature and the Bochner Laplacian, as well as their corresponding flows.

Until quite recently, Ricci curvature has been just one of the various curvature notions in Riemannian Geometry, best known for its role in Einstein's field equations. However, this has changed drastically due to the spectacular mathematical applications in G. Perelman’s far reaching work on the Ricci flow and the Geometrization Conjecture \cite{Per1,Per2}.  
 
While started far earlier \cite{Stone}, the search for the discretization of this notion has gained fast momentum, spurred by Perelman's groundbreaking results. One notable approach is that of Chow and Luo \cite{CL}, based on circle packings. It 
has been successfully applied in Graphics, Medical Imaging and Communication Networks (see e.g. \cite{YJLG}) giving rise to many practical applications. A radically different approach was adopted by Ollivier \cite{Ol1,Ol2}, that proved to be excellently suited for modeling Complex Networks, both from the theoretical  \cite{BJL,JL,LR}, as well as applied \cite{Allen1,Allen2,NLGGS,Shiping} viewpoints. 
%Furthermore, as a tools for rending further related theoretic results, it has the promise of further applicative potential .
% \add[MW]{References not relevant.}

Recently, another discretization of Ricci curvature for complex networks based on Forman's theoretical work \cite{Fo} has been introduced \cite{SMJSS,SJSS,WJS}. While based on different ideas than those residing at the base of Ollivier's discretization, Forman's version is far simpler to implement - an advantage that was emphasized and exploited in the papers mentioned above. Furthermore, there are strong theoretical indications that the two approaches are closely correlated, suggesting to substitute Ollivier's curvature with Forman's for efficient computation in practical applications.

The formalism discussed in the present article naturally applies to both static and dynamic networks. We discuss the duality of curvature and flow in a more general, theoretic setting and its exploitation in Network Analysis. Both concepts are used to investigate model and real-world networks, including web graphs (Google), social networks (Facebook) and biological networks (gene networks, BioGrid), in terms of the Ricci-formalism. The results are discussed in comparison to node-degree based network analysis and standard models as classifiers of complex networks representing real-world systems.
In the second part of the article, we focus on characterizing dynamic networks with the help of the Ricci flow in an attempt to detect and describe the topological properties of the underlying dynamic effects. We evaluate our results in comparison to the Laplacian flow, a node-based method related to both the Forman curvature and the Bochner Laplacian.

We conclude with possible applications of the discussed methods in data mining with an emphasis on the analysis of dynamic effects in complex systems.

%\add[MW]{Add note: In the final section we compare the edge-based Forman-Ricci flow with the node-based Laplacian flow.}

%\add[MW]{This is just an outline/ summary of ideas.}
%\add[MW]{It would be great, if you could send me a list of things you would like to adress in the introduction or maybe you could sketch something (plain text/ Word is fine).}

%\begin{footnotesize}
%\begin{itemize}
%\item goals/ questions to adress in data characterization
%\item importance: wide variety of data sets for which methods can be applied (social nets, p2p, biological - neuro + genetics)
%\item dualism: Curvature and Laplacian \add[MW]{Emil}
%\item review work of Jost group, Forman, Bochner, Olivier ... 
%\item data science applications for our methods: change detection + classification
%\end{itemize}
%\end{footnotesize}

\section{Forman-Ricci curvature and the Bochner Laplacian}

In this section we will give a mathematical motivation and rigorous introduction of the Ricci-formalism for networks. We will discuss essential properties of Forman's Ricci curvature and the resulting important consequences for complex networks before moving on to possible practical applications in Network Analysis in the next sections. 

%Since the definition of Forman's curvature also prescribes some of its essential properties that have important consequences in certain significant applications, we shall detail somewhat its definition, and not satisfy ourselves with bringing the technical defining formula of the Forman-Ricci curvature.

%\begin{footnotesize}
%\begin{itemize}
%\item Curvature (Forman-Ricci)
%\item Laplacian: Combinatorial, Bochner, other definitions?
%\item dualism: curvature/ flow and Laplacian
%\item Gauss-Bonnet for normalized Laplacian (Ethan Bloch)
%\end{itemize}
%\end{footnotesize}

\subsection{Ricci curvature}
In the classical, geometric sense, Ricci curvature measures the deviation of a manifold from being locally Euclidean % operates directionally ,
in various tangential directions. More precisely, it quantifies both divergence (of geodesics) and (volume) growth in the second term of the $(n-1)$-dimensional volume $\Omega(\varepsilon)$ through controlling the growth of the measured angles, as quantified by the following classical formula:

	\begin{equation}
	\Omega(\varepsilon)={\rm Vol}\big(\varpi(\alpha)\big) = d\alpha\,\varepsilon^{n-1}\left(1 - \frac{{\rm Ric}({\bf v})}{3}\varepsilon^2 +
	o(\varepsilon^2)\right)\,.
	\end{equation}
where $d\alpha$ denotes the $n$-dimensional solid angle in the direction of the vector ${\bf v}
\in T_p(M^n)$
and  $\varpi(\alpha)$ represents the $(n-1)$-volume generated by geodesics of length $\varepsilon$ in $d\alpha$. Forman's discretization of the Ricci curvature specifically captures the volume growth and transfers the formalism to networks.

%\marginpar{\color{RedClr} To-do: Modify figure to avoid copyright-issues.}
%\marginpar{\color{RedClr} \tiny \bf I can include a picture here, but I am not sure I want to repeat to much what I already wrote in previous papers... \\ \color{BlueClr} I think we could use a sketch here, I personally always find a figure for geometric arguments very helpful. I'll sketch something and will send it to you. } 

\begin{figure}[H]
	\centering
	% \captionsetup{width=.9\linewidth}
	\includegraphics[width=.5\linewidth,natwidth=1518 ,natheight= 949]{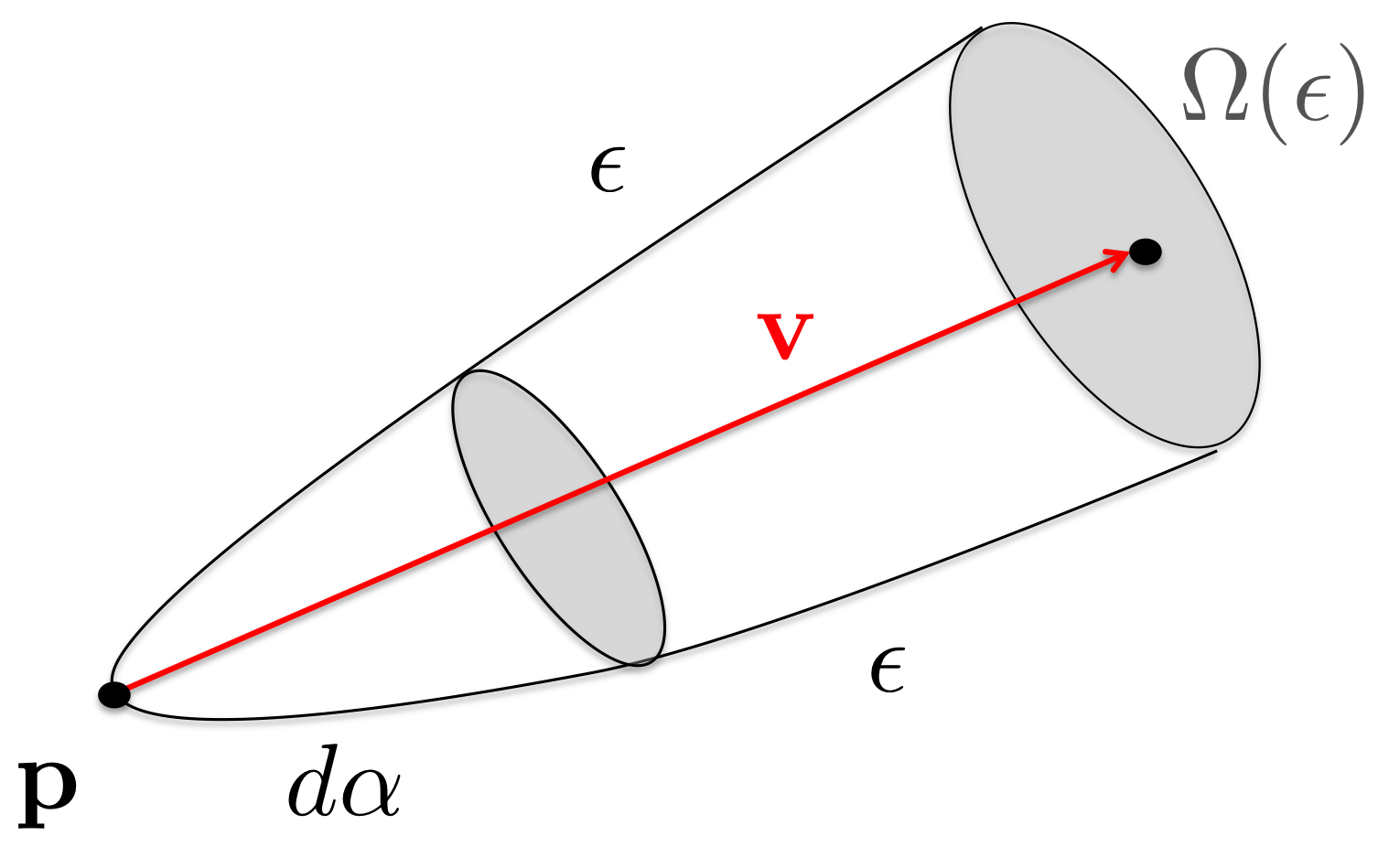} 
	\caption[Ricci curvature as a deviation of a manifold from being locally Euclidean]{Ricci curvature as deviation of a manifold from being locally Euclidean % operates directionally ,
		in various tangential directions (after Berger \cite{Be}). Here
		$d\alpha$ denotes the solid angle in the direction
		of the vector ${\bf v}$, and $\Omega$ the volume
		generated by the geodesics of length $\epsilon$ in $d\alpha$.}
	\label{fig_sim}
\end{figure}

%\marginpar{\color{RedClr} To-do: Modify figure to avoid copyright-issues.}
%\marginpar{\color{RedClr} \tiny \bf  I have a ready made picture, but I used it before, so a small modification would be nice... I might have a colored figure somewhere, if I find it, then that will do. \\ \color{BlueClr} This looks good. Yes, we need at least a small modification to avoid copyright issues. Color might do this - but we need to check, if the journal allows colored pictures and if they charge extra publication fees (I know of at least one (also Oxford) journal who does).}

\noindent Technically, Ricci curvature represents an average of {\em sectional
curvatures}. 
%\marginpar{\color{RedClr} \tiny \bf I can add a formula here, but I am not sure is necessary - we emphasize different aspects here. \\ \color{BlueClr} I think text + figure are pretty clear, maybe better not make it more technical at this point.}
As an analogy to the classical {\it mean curvature} of surfaces, this \textit{averaging property} is further emphasized by
the fact that Ricci curvature acts as the Laplacian of
the metric $g$ \cite{Be}.
Notably, in dimension $n=2$, i.e. in the case most relevant for classical Image Processing and
related fields, Ricci curvature reduces to sectional (and
scalar) curvature in the form of the classical case \cite{Be}.
%{\it Gauss curvature}. \add[MW]{reference?}

\subsection{The Bochner Laplacian}
Before proceeding to the more technical aspects, we note that Forman's definition has a quite general discrete setting. It is based on an abstraction of a classical formula known from Differential Geometry and Geometric Analysis as the so called {\it Bochner-Weitzenb\"{o}ck formula} (see, e.g. \cite{J}), for {\it weighted $CW$ cell complexes}, an abstraction of both polygonal meshes and weighted graphs. This formula relates curvature to the classical (Riemannian) Laplace operator. In consequence, and as a byproduct of its very definition, Forman's Ricci curvature comes coupled with a fitting Laplacian (or, in fact, two Laplacians, as we shall see below). The parallelism of the Forman-curvature and the Bochner Laplacian, also reflected in the corresponding flows, has important implications on the transfer of theoretical results and possible applications
onto various fields of Network Analysis.
%through analogies in related applied fields of network analysis, such as Imaging and Graphics (see, e.g. \cite{Xu} and the references therein).

In its best known form, for functions, the Bochner-Weitzenb\"{o}ck formula is written as:
\begin{equation} \label{eq:BW1}
-\frac{1}{2}\Delta\left(||df||^2\right) = ||{\rm Hess}f||^2 - <df,\Delta df> + {\rm Ric}(df,df)\,.
\end{equation}
Here ${\rm Hess}f$ denotes the Hessian of $f$: ${\rm Hess}f = \nabla df = \nabla^2f$ and $<\cdot,\cdot>$ as usual, the inner product.
A proof can be found, e.g., in \cite{J}\footnote{Besides proofing the above statement, \cite{J} emphasizes the role of Ricci curvature in the formula of the Jacobian determinant of the exponential map, thus underlining the role of Ricci curvature as a measure of growth -- see also the observation above regarding this very aspect of Forman's discretization.}. However, there is no natural or general applicable interpretation of this form of the Bochner formula. Therefore, to generalize the notion of Ricci curvature to allow for a description of weighted cell complexes, one starts
from the following form of the {\it Bochner-Weitzenb\"{o}ck} formula (see, e.g. \cite{Be,J}) for the {\it
Riemann-Laplace operator} $\Box_p$ on $p$-forms on (compact) Riemannian manifolds:
\begin{equation} \label{saucan-eqn:1}
\Box_p = dd^* + d^*d = \nabla_p^*\nabla_p  + {\rm Curv}(R)\,,
\end{equation}
where $\nabla_p^*\nabla_p$ is the {\it Bochner} (or {\it rough}) {\it Laplacian} and ${\rm Curv}(R)$ an %complicated
expression of the {\it curvature tensor} with linear coefficients  %(i.e. in the (second order?!) derivatives of the metric).
where $\nabla_p$ denotes the {\it covariant derivative} operator. % (on $\Omega^p(M)$).
For cell-complexes, one of course cannot expect the existence of such differentiable
operators. However, a {\it formal} differential exists: In our
combinatorial context (the operator) ``$d$'' being replaced by
``$\partial$'' -- the boundary operator of the cellular chain
complex (see \cite{KMM}), we have
%\add[MW]{reference R not listed in the bibliography}),
%
\[0 \rightarrow C_n(M,\mathbb{R}) \stackrel{\partial}{\rightarrow} C_{n-1}(M,\mathbb{R}) \stackrel{\partial}{\rightarrow}
\cdots \stackrel{\partial}{\rightarrow} C_{0}(M,\mathbb{R}) \rightarrow 0\,,\]
where cells are analogues of forms in the classical (i.e. Riemannian) setting.
The following definition of the Bochner Laplacian becomes now
natural:

\begin{equation} \label{saucan-eqn:2}
\Box_p = \partial\partial^* + \partial^*\partial: C_p(M,\mathbb{R}) \rightarrow C_p(M,\mathbb{R})\,,
\end{equation}
where  %$\partial^*$
$\partial^*:C_p(M,\mathbb{R}) \rightarrow C_{p+1}(M,\mathbb{R})$ is the {\it adjoint} (or {\it coboundary})
operator of $\partial$, defined by:
\(<\partial_{p+1}c_{p+1},c_p>$ $=
<c_{p+1},\partial_p^*c_p>_{p+1}\,,\)
where $<\cdot,\cdot> = <\cdot,\cdot>_p$ is a (positive definite)
inner product on $C_p(M,\mathbb{R})$, i.e. satisfying: (i)
$<\alpha,\beta> = 0, \forall \alpha \neq \beta$ and (ii)
$<\alpha,\alpha> = w_\alpha > 0$ -- the weight of cell $\alpha$.
%%
%%\marginpar{\color{RedClr} \tiny \bf This might be too much! We can move part (or most) of it to an Appendix. \\ \color{BlueClr} I would keep it, since it fits well into the context and we need the theory, if we include a comparison of the Bochner-Laplacian with the Ricci-flow in the section below.}
%%----------------------------------------------------------------------

Forman \cite{Fo} (see also the much earlier \cite{Eck}) shows that an analogue of the Bochner-Weitzenb\"{o}ck formula holds in this setting, i.e. that
there exists a canonical decomposition of the form:
\begin{equation} \label{saucan-eqn:3}
\Box_p = B_p + F_p\,,
\end{equation}
where $B_p$ is a {\it non-negative operator} and $F_p$ is a diagonal matrix. $\Box_p$, $B_p$ and $F_p$ are called,
in analogy with the classical Bochner-Weitzenb\"{o}ck formula, the {\it combinatorial Riemann-Laplace operator},  the  \textit{combinatorial Bochner} (or \textit{rough})  \textit{Laplacian}, and the {\it combinatorial curvature function}, respectively. 

If $\alpha = \alpha^p$ is a $p$-dimensional cell (or $p$-cell, for short), then we can define the {\it
curvature function} $F_p: C_p \rightarrow C_p$
\begin{eqnarray}
\mathcal{F}_p = <F_p(\alpha),\alpha>,
\end{eqnarray}
as a linear function on $p$-chains. For dimension $p=1$ we obtain, by analogy with the classical case, the
following definition of discrete (weighted) {\it Forman-Ricci curvature} on $\alpha = \alpha^1$, i.e. 1-cells (edges):
\begin{eqnarray}
\label{saucan-def:Ricci}
{\rm Ric_F}(\alpha)= \mathcal{F}_1(\alpha).
\end{eqnarray}
\noindent The formalism does not restrict the choice of the weight function, i.e. it allows for general weights making the Forman-Ricci curvature extremely versatile. A handy choice are the so called {\it standard weights} \cite{Fo}, %Theorem 2.5 and Theorem 3.9)
that generalize the intuitive geometric notions of length, area and volume:%\\\\
\begin{equation} \label{eq:scaling}
w(\alpha^p) = w_1\cdot w_2^p\,.
\end{equation}
\noindent Note that the {\em combinatorial weights} $w_\alpha \equiv 1$ represent a
set of standard weights, with $w_1 = w_2 = 1$. 

Using the properties of standard weights and their relationship with generic weights, we obtain the following formula for polyhedral (and in fact much more general) complexes, endowed with any set of (positive) weights:

\begin{equation} \label{eq:Forman-ch1}
\hspace*{-0.75cm}
\mathcal{F}(\alpha^p) = w(\alpha^p)\Big[\Big(\sum_{\beta^{p+1}>\alpha^p}\frac{w(\alpha^p)}{w(\beta^{p+1})}\;
%\]
%%
%\[
+ \sum_{\gamma^{p-1}<\alpha^p}\frac{w(\gamma^{p-1})}{w(\alpha^p)}\Big)\; 
\end{equation}
\[
\hspace*{3.2cm}
-\sum_{\alpha_1^p\parallel \alpha^p, \alpha_1^p \neq \alpha^p}\Big|\sum_{\substack{\beta^{p+1}>\alpha_1^p \\ \beta^{p+1}>\alpha^p}}\frac{\sqrt{w(\alpha^p)w(\alpha_1^p)}}{w(\beta^{p+1})}
%\]
%
%\hspace*{-1.5cm}
- \sum_{\substack{\gamma^{p-1}<\alpha_1^p \\ \gamma^{p-1}<\alpha^p}}\frac{w(\gamma^{p-1})}{\sqrt{w(\alpha^p)w(\alpha_1^p)}}\Big|\:\;\Big]\,;
%\end{equation}
\]
\noindent with $\alpha < \beta$ meaning that $\alpha$ is a face of $\beta$, and
$\alpha_1 \parallel \alpha_2$ signifying the
simplices $\alpha_1$ and $\alpha_2$ to be {\it parallel}.

\begin{figure}[H]
	\centering
%  \captionsetup{width=.9\linewidth}
	\includegraphics[width=.5\linewidth,natwidth=1009 ,natheight= 1014]{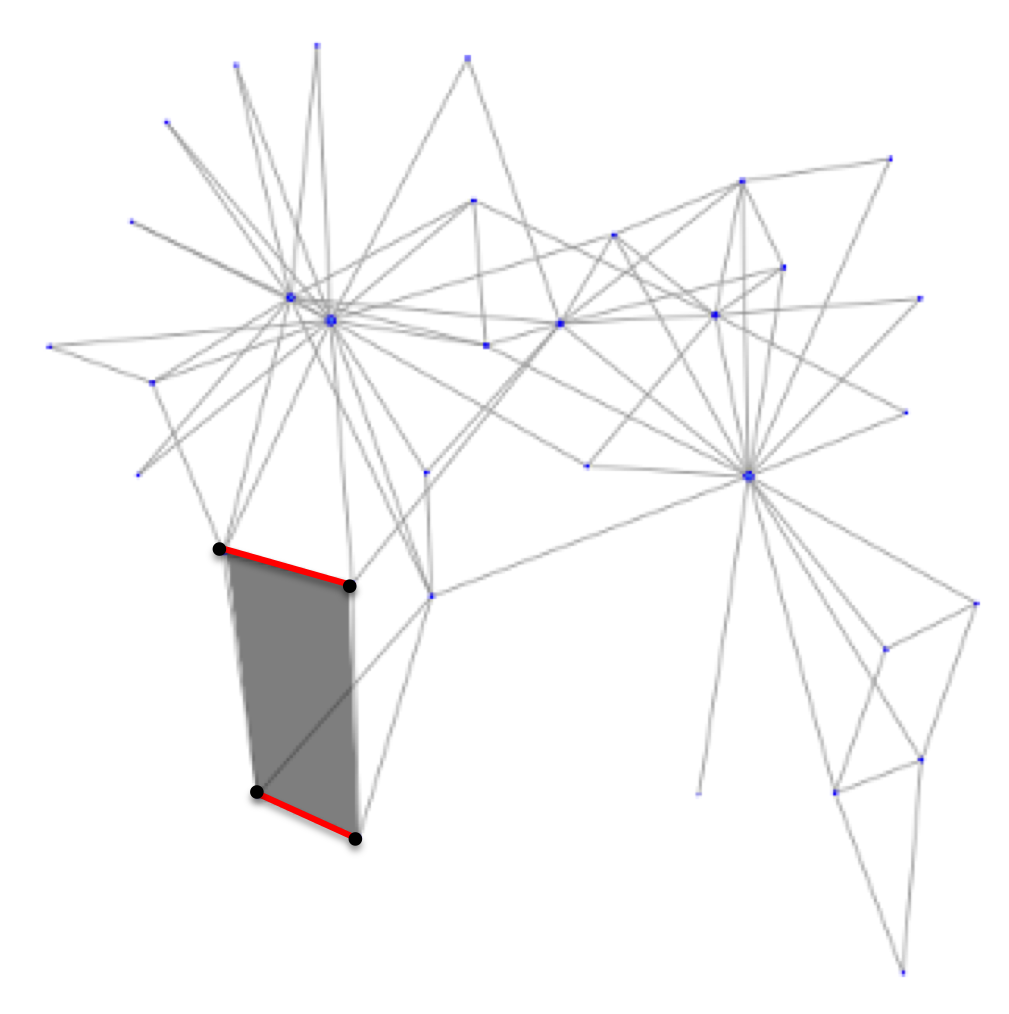} 
	\caption[Parallel cells]{{\small Social network \cite{karate} with highlighted cells. Parallel cells (edges: red), with distinct children (nodes: black) and a common parent (face: gray).}}
	\label{fig:parallel-cells}
\end{figure}
%\add[ES]{This caption wasn't correct: The whole "ensamble" is a complex (a type of space).}

%\marginpar{\color{RedClr} \bf \tiny Should we again insert a picture? \\ \color{BlueClr} Here, again, I think a picture would be helpful to understand the respective paragraph. I can sketch something for this.}

\noindent Parallel cells are precisely those that either have a common ``\textit{parent}" (adjacent higher dimensional face) or a common ``\textit{child}" (adjacent lower dimensional face), but not those that have both a common parent and common child (see Fig. (\ref{fig:parallel-cells})). Together with the formula above, the combinatorial Riemann-Laplace operator (see \cite{Fo}) is also
obtained, when we consider the Laplacian written in not necessarily symmetric form (although, with a suitable choice of an inner product, the Laplacian is a symmetric operator, see \cite{HJ}), as:
\begin{equation}  \label{saucan-eqn:Lap}
\hspace*{1.3cm}
%\[\hspace*{-0.2cm}
\Box_p(\alpha_1^p,\alpha_2^p) =
\sum_{\substack{\beta^{p+1}>\alpha_1^p \\ \beta^{p+1}>\alpha_2^p}}\epsilon_{\alpha_1,\alpha_2,\beta}\frac{\sqrt{w(\alpha_1^p)w(\alpha_2^p)}}{w(\beta^{p+1})}\:
%\]
 %
%
%
+  \sum_{\substack{\gamma^{p-1}<\alpha_1^p \\ \gamma^{p-1}<\alpha_2^p}}\epsilon_{\alpha_1,\alpha_2,\gamma}\frac{w(\gamma^{p-1})}{\sqrt{w(\alpha_1^p)w(\alpha_2^p)}}\,;
\end{equation}
where $\epsilon_{\alpha_1,\alpha_2,\beta},\epsilon_{\alpha_1,\alpha_2,\gamma} \in \{-1,+1\}$ are the relative orientations of the cells.

\begin{rem}
One might object that the formulas above hold only for the case of standard weights, whereas weights associated to intrinsic information in the real data need not necessarily satisfy such dimensionality scaling -- like the one displayed by area and volume with respect to the length, as encapsulated by condition (\ref {eq:scaling}). In particular, while this scaling holds for polygonal and polyhedral meshes, it does not necessarily hold for networks. 
However, an important result of Forman (\cite{Fo}, Theorem 2.5) consist precisely in showing that, for an open, dense set of weights, the essential property required for the formulas in question to hold is satisfied. Therefore, while the given weights might not satisfy the required property themselves, there exist weights arbitrarily close to them that do. Hence they can replace, formally, the given ones, without any perceivable computational error.
\end{rem}

%----------------------------

\subsection{The case of networks}

In 1-dimensional regular $CW$-complexes, such as networks or locally finite graphs, there are no faces of dimension higher than one. It follows for these cases including the complex networks considered here, that edges have no ``parents'', but only ``children'' (nodes). This observation greatly simplifies (\ref{eq:Forman-ch1}), since it causes half of the terms to vanish. The case in which we do have these higher (and lower) order structures will be covered in a forthcoming article by the authors.

By further taking $p = 1$, we obtain the Forman-Ricci curvature for networks:

\begin{equation} \label{FormanRicciEdge}
{\rm Ric_F}(e) = \omega (e) \left( \frac{\omega (v_1)}{\omega (e)} +  \frac{\omega (v_2)}{\omega (e)}  - \sum_{\substack{e_{v_1}\ \sim\ e \\ \ e_{v_2}\ \sim\ e}} \left[\frac{\omega (v_1)}{\sqrt{\omega (e) \omega (e_{v_1})}} + \frac{\omega (v_2)}{\sqrt{\omega (e) \omega (e_{v_2})}} \right] \right)\,.
\end{equation}

\noindent Analogously to (2.8), the Bochner Laplacian simplifies, for $p=1$, to

\begin{eqnarray}
\label{weber:FormanLap}
\Box_1 (e_1 , e_2) = \sum_{\substack{e_1 \sim v \\ e_2 \sim v}} \frac{\omega (v)}{\sqrt{\omega (e_1) \omega (e_2)}},
\end{eqnarray}

\noindent due to vanishing terms related to the non-occurrence of higher order faces. In section 4, we will use (\ref{FormanRicciEdge}) and (\ref{weber:FormanLap}) to characterize the flow on networks.

As underlined above, the main advantage of Forman-Ricci curvature as a network characteristic resides in the fact that it is an edge based notion. However, it can be preferable to use a node based version. In that case, the Forman-Ricci curvature of a node can be defined as the sum of the curvatures of all edges incident to that node:
\begin{equation}
\label{UnnormalizedFormanNode}
\mathbf{F}(v) = \sum_{e_v\ \sim\ v} \mathbf{F}(e_v) \,.
\end{equation}

\noindent Directed (oriented) networks are essential for modeling a variety of phenomena, such as those included in the experimental part of the present paper (see Section 3).
However, in Forman's original work, the weights (as generalizations of length, area, etc.) were taken to be positive. Hence, the Forman-Ricci curvature is not directly applicable for non-orientable surfaces (or, more generally, manifolds/ cell complexes). However, one can adapt its notion from undirected to directed graphs by rearranging the terms in the definition of the Forman curvature for an edge $e$ (Eq. \ref{FormanRicciEdge}), and write the contributions of the two adjacent nodes $v_1$ and $v_2$ separately (see also \cite{SJSS}) as
\begin{equation}
\label{FormanRicciEdgeAlternate}
\mathbf{F}(e) = w_e \left( \frac{w_{v_1}}{w_e}   - \sum_{e_{v_1}\ \sim\ e} \frac{w_{v_1}}{\sqrt{w_e w_{e_{v_1} }}} \right)  + w_e \left( \frac{w_{v_2}}{w_e}   - \sum_{\ e_{v_2}\ \sim\ e} \frac{w_{v_2}}{\sqrt{w_e w_{e_{v_2} }}}  \right) \; .
\end{equation}
Note, that this naturally defines the curvature of a directed edge by only using the term involving its initial node, or alternatively its terminal node. 

This definition can be, in turn, easily adapted to nodes: 
Given a node $v$, let us denote the set of \textit{incoming} and \textit{outgoing} edges of $v$ by $E_{I,v}$ and $E_{O,v}$, respectively. Then, one can define the \textit{In Forman curvature} $\rm{Ric_{F, I}}(v)$ and the \textit{Out Forman curvature} $\rm{Ric_{F,O}}(v)$ as follows:
\begin{equation}
\label{FormanRicciNodeIn}
{\rm Ric_{F,I}}(v) = {\sum_{e \in E_{I,v}} {\rm Ric_{F,I}}(e)}\,;
\end{equation}
\begin{equation}
\label{FormanRicciNodeOut}
{\rm Ric_{F,O}}(v) = {\sum_{e \in E_{O,v}} {\rm Ric_{F,O}}(e)}\,;
\end{equation}
where the summations are taken over only the incoming and outgoing edges, respectively. Moreover, one can obtain the total amount of flow through a node $v$ as follows:
\begin{equation}
\label{FormanRicciNodeFlow}
{\rm Ric_{F, I/O}}(v) = {\rm Ric_{F,I}}(v) - {\rm Ric_{F,O}}(v)\,.
\end{equation}

\noindent Note that we extend formula (2.8) and define the Ricci curvature of a directed edge, by using just one of its two nodes, i.e. either its ``head'' or its ``tail''. Here we measure the incoming edges at the head and the outgoing ones at the tail where directions are defined from head to tail. While there are other possible choices, we preferred this convention, as it is the most natural for modeling a variety the networks, in particular those analyzed in the present article.

\section{Application on Complex Networks}
%\add[MW]{This is just a rough draft.}

\noindent In the previous section, the Forman-Ricci curvature was rigorously introduced as a network property. Now we want to investigate its applicability as a characteristic for real-world complex networks.

As already noted above, common network characteristics are usually node-based features, i.e. greatly dependent on node-degrees. While there are a great number of proposed approaches, typically, the node degree distribution, the average path length between any two nodes and the clustering coefficient \cite{newman1} are used as ``gold standards" for evaluating and classifying networks (a notable exception to this common approach can be found in [Bai et al., 2016]). All of these properties are highly dependent on node degrees, over-emphasizing the structural importance of nodes over edges. 

This exaggerated centrality of node degree-based features holds also for the three most established model networks  (respectively, Erdős-Rényi \cite{er1, er2}, Watts-Strogatz \cite{ws} and Albert-Barabási \cite{ab}), that are widely used as exemplary models for real-world networks. Their common focus on node-based network properties has been both apprised \cite{stanley,jeong2} -- for their simplicity -- and criticized  \cite{arita_metabolic_2004, marcotte} -- for significant qualitative deviations from real world networks.

In contrast, in this article, we examine the applicability of the edge-based Forman-Ricci curvature as a characteristic for complex networks. Many real-world networks are naturally weighted and other than in the case of standard weights imposed onto unweighted networks, we assume edge weights to be independent of node degrees. However, it has been shown that many weighted real-world networks, especially biological ones, have a node degree bias (see, e.g. \cite{egad} and the references therein). Even if curvature cannot be assumed independent of node degrees, it balances their dominance in network analysis by taking other important information - namely the edge weights - into account. We believe, that an extended network analysis that goes beyond the classic ``gold standard" of network properties to include the curvature and possibly the spectrum of the graph Laplacian \cite{jost1,jost2}, gives a more complete picture and provides better insight into the topology of complex networks.

For the exploratory investigation of complex networks in this section, we follow current developments and areas of major interest in the network sciences and related fields and choose the following exemplary data sets:

\begin{enumerate}

\item \textbf{Webgraphs} \\
Following the increasing impact of the World Wide Web on today's world, the topology of the internet has attracted major interest in recent years. We investigate a Google web graph from \cite{google,konect}, where nodes represent web pages and edges the respective hyperlinks between them.

\item \textbf{Social graphs} \\
The rapid rise of social networks like Facebook, LinkedIn and Twitter have provided the social sciences with fundamentally new ways to study human interactions and social structures. Research now focuses on questions closely related to the topology of social graphs that represent the underlying complex systems. In this section, we study a Facebook friendship graph from \cite{facebook,snap}, in which nodes denote users and edges represent the connections (friendships) between them.

\item \textbf{Biological networks} \\
In recent years, data mining and network analysis have revolutionized a great number of disciplines in the biological sciences. For instance, networks built from co-expression data or protein-protein interactions in Genomics as well as brain networks in Neurosciences are widely used data representations that have given rise to fundamentally new methods and scientific approaches. Besides novel insights, these methods also demonstrate a need to understand the topology of networks for constructing null models and reliable means to account for noise biases that are an unavoidable feature of systems modeled from experimental data (see, e.g. \cite{egad} and the references therein). Here, we investigate a gene-interaction network built from BioGrid \cite{biogrid}, where genes are represented by nodes and their associations and commonalities by the respective edges.

\end{enumerate}

%We should note here that it is quite straightforward to extend formula (2.6) and define the Ricci curvature of a directed edge, by using just one of its two nodes, i.e. either its ``head'' or its ``tail''. Here we used the convention  of measuring the incoming edges at the head and the outgoing ones at the tail. While this choice is, by no means, the only possible one, we preferred it, as being the most natural for modeling a variety the networks, in particular those analyzed in the sequel.

%\begin{footnotesize}
%\begin{itemize}
%\item Curvature distribution and curvature maps as network characteristics
%\item use as classifier, together with classic characteristics and Laplacian graph spectrum as introduced in \cite{jost1, jost2}
%\item data analysis: social graph (facebook), web graph (google), biological network (genetic - biogrid human); one panel each (curvature map + distribution, Laplacians)
%\item comparison with standard models: ER, WS, AB (?)
%\item \add[MW]{Emil:} Clustering with Forman-curvature
%        \add[ES]{Since this is only an idea/proposal, shouldn't we better leave it for the Future Work section?}
%\end{itemize}
%\end{footnotesize}

\subsection{Forman-Ricci curvature as a network characteristic}

\subsection*{\textbf{Complex Networks}} 
\noindent In this article, we consider network graphs
\begin{align*}
G = \lbrace V, E \rbrace \; ,
\end{align*}
where $V(G)$ represents the set of nodes (or vertices) and $E(G)$ the edges connecting them according to known associations, interactions or commonalities. Those can be inferred from a given data set, empirical information or previous knowledge.

We impose a weighting scheme on the nodes 
\begin{eqnarray}
\omega : V(G) \mapsto [0,1] \; ,
\end{eqnarray}
that can reflect additional knowledge about the nodes (such as hierarchical positions in social networks, or the number of monthly visitors of a webpage) or - if no previous information on the data points,  other than their associations is given - we construct them combinatorial, based on node degrees:
\begin{eqnarray}
\omega (v) = \frac{1}{{\rm deg}(v)} \sum_{u \sim v} {\deg (u)} \; .
\end{eqnarray}
%\add[ES]{The idea is cleary explained. I noticed that  you removed the part (formula) with $dist \leq 6$, which I find now less clear in the conference paper. I think we shoould take anothet look at this again, to make sure everything is alright. (Also, a reviwer might ask for more motivation why we are doing this.) }

\noindent To account for the strength of these associations, we define a normalized weighting scheme
\begin{eqnarray}
\gamma : E(G) \mapsto [0,1] \; .
\end{eqnarray}
The weights can be imposed directly from the given data (e.g. in correlation networks, as the correlation between two data points across a set of empirical data) or derived from the weights of the nodes they connect:
\begin{equation}
\gamma (e_{ij}) = {\rm sign}(e_{ij}) \sqrt{\omega(v_i)^2 + \omega(v_j)^2} \; ,
\label{eq:weight_e}
\end{equation}

\noindent with
\begin{align}
{\rm sign}(e_{ij}) = \begin{cases} 1,  & i \leq j \\ 
-1, & \mbox{else}		
\end{cases} \; .
\end{align}

\noindent The weighting scheme Eq.(\ref{eq:weight_e}) is motivated by the geometric analogy of the distance between two points and can be extended to a more generalized weighting scheme for cellular complexes. % ... as adressed in the second part of this thesis.
%
%\add[MW]{I modified the default weighting scheme of the edges. This mainly due to a better geometric motivation in preparation of extending the weighting schemes to higher order faces in the following paper. Since they'll together be the thesis, the weighting scheme should be consistent.}
%
Using edge and node weights, we can now determine the Forman curvature distribution from (\ref{FormanRicciEdge}) (or, in the directed case, from (\ref{FormanRicciNodeIn}) and (\ref{FormanRicciNodeOut})). For evaluation of the distribution, we use both histograms and curvature maps, the later ones containing through their spatial components additional information on directionality and community structure.

\subsection*{\textbf{A curvature-based distance measure for graphs}}

\noindent Using the distribution of the Forman-Ricci curvature we want to introduce a tool for determining the distance between network graphs. We introduce a curvature-based pseudo-distance as a measure of the dissimilarity of pairs of graphs that gives rise to a curvature-based classification scheme.

Firstly, we estimate the density of the distribution through a (normalized) kernel density estimator
\begin{align}
f_G : {\rm Ric_F} (G) \rightarrow \mathbb{R} \; ; \\
f_G (e) = \frac{1}{|\hat{e} |} \sum_{\hat{e} \sim e} K \left( {\rm Ric_F}(e) - {\rm Ric_F} \left( \hat{e} \right) \right) \; ,
\end{align}

\noindent with a kernel function $K$. Here, $\hat{e}$ are edges parallel to $e$. To characterize the distance between a pair of network graphs $G_1$ and $G_2$, we want to calculate the distance between the respective density estimates $f_{G_1}$ and $f_{G_1}$. A distance measure for distributions is given by the \textit{Wasserstein}-distance \cite{gibbs-su}
%
%\begin{eqnarray}
%d_W = \int_{x=-\infty}^{\infty} \lvert \int_{y=-\infty}^{x} \left( f_{G_1} (y) -  f_{G_2} (y) \right) dy \rvert dx
%\end{eqnarray}
\begin{align}
d_W ^p &= \inf \; \mathbb{E} \left( d\left( f_{G_1}, f_{G_2}	\right)^{1/p}	\right) \\
&= \left( \int_{\mathbb{R}} \vert f_{G_1}(x) - f_{G_2}(x)	\vert^p dx \right)^{1/p} \; .
\end{align}
Here, we will use a special form of the Wasserstein-1 metric. For this, we formulate the graph distance as an optimal transport problem and use the \textit{earth mover's distance}, i.e. $d_W^1$, that can be computed by the \textit{Hungarian algorithm}  \cite{rubner}. We define clusters by splitting ${\rm supp} f_{G_{1,2}}$ into $k_{1,2}$ \textit{bins}, i.e. curvature-intervals $\lbrace x_{1,2}^i\rbrace_{i=1, ... , k}$ such that
\begin{eqnarray}
\bigcup_{i=1}^k x_{1,2}^i = {\rm supp} f_{G_{1,2}} \; .
\end{eqnarray}

\noindent Each bin defines a cluster $P_{G_{1,2}}=\lbrace P_{1,2}^1, ... , P_{1,2}^{k_1,k_2} \rbrace$ where $P_{1,2}^i$ is characterized by a representative $p_{1,2}^i$ with weight $\omega (p_{1,2}^i )$, i.e.
\begin{align}
p_{1,2}^i = \langle	 x_{1,2}^i \rangle \; ; \\
\omega(p_{1,2}^i) = f_{G_1,G_2} \left( \langle x_{1,2}^i \rangle \right) \; ;\\
P_{1,2}^i=\left( p_{1,2}^i, \omega(p_{1,2}^i)	\right) \; .
\end{align}

\noindent The pairwise distances between the binned clusters of both graphs are the entries of the so-called \textit{ground distance matrix}
\begin{eqnarray}
D= \lbrace	d_{ij} \rbrace_{\substack{i=1, ... , k_1 \\ j= 1, ... , k_2}} = {\rm dist} \left(	p_{1}^i, p_{2}^j\right) \; ,
\end{eqnarray}

\noindent The distance between the graphs can also be viewed as the (finite) minimal number of modifications (i.e. adding or removing edges or nodes) to transform $G_1$ into $G_2$. In our setting, this corresponds to an optimal transport problem between the sets of clusters $P_{G_1}$ and $P_{G_2}$. We introduce the \textit{transportation flow}
\begin{eqnarray}
F= \lbrace	f_{ij} \rbrace_{\substack{i=1, ... , k_1 \\ j= 1, ... , k_2}} \; ,
\end{eqnarray}
for characterizing the transformation of $P_{G_1}$ to $P_{G_2}$. To find the optimal transport F, we minimize the \textit{transport cost}
\begin{eqnarray}
W\left(	P_{G_1}, P_{G_2}, F \right)= \sum_{i=1}^{k_1} \sum_{j=1}^{k_2} f_{ij} \cdot d_{ij} \; .
\end{eqnarray}
The optimization is implemented by, e.g. the Hungarian algorithm \cite{rubner}. With the resulting optimal $F$, we can calculate the distance between $G_1$ and $G_2$ by the \textit{discrete earth mover's distance}
\begin{eqnarray}
EMD \left(	P_{G_1}, P_{G_2} \right) = d_W^1 = \frac{\sum_{i=1}^{k_1} \sum_{j=1}^{k_2} f_{ij} d_{ij}}{\sum_{i=1}^{k_1} \sum_{j=1}^{k_2} f_{ij}} \; .
\end{eqnarray}

\subsection{Computational Data Analysis}
 
%\begin{itemize}
%\item evaluation of results
%\item comparison with standard models
%\item compute distances of distributions (?)
%\end{itemize}

\subsection*{\textbf{Distribution and Similarity}}

\begin{figure}[htb]
	\centering
%	\captionsetup{width=.9\linewidth}
	\includegraphics[width=\linewidth,natwidth=2127,natheight= 1226]{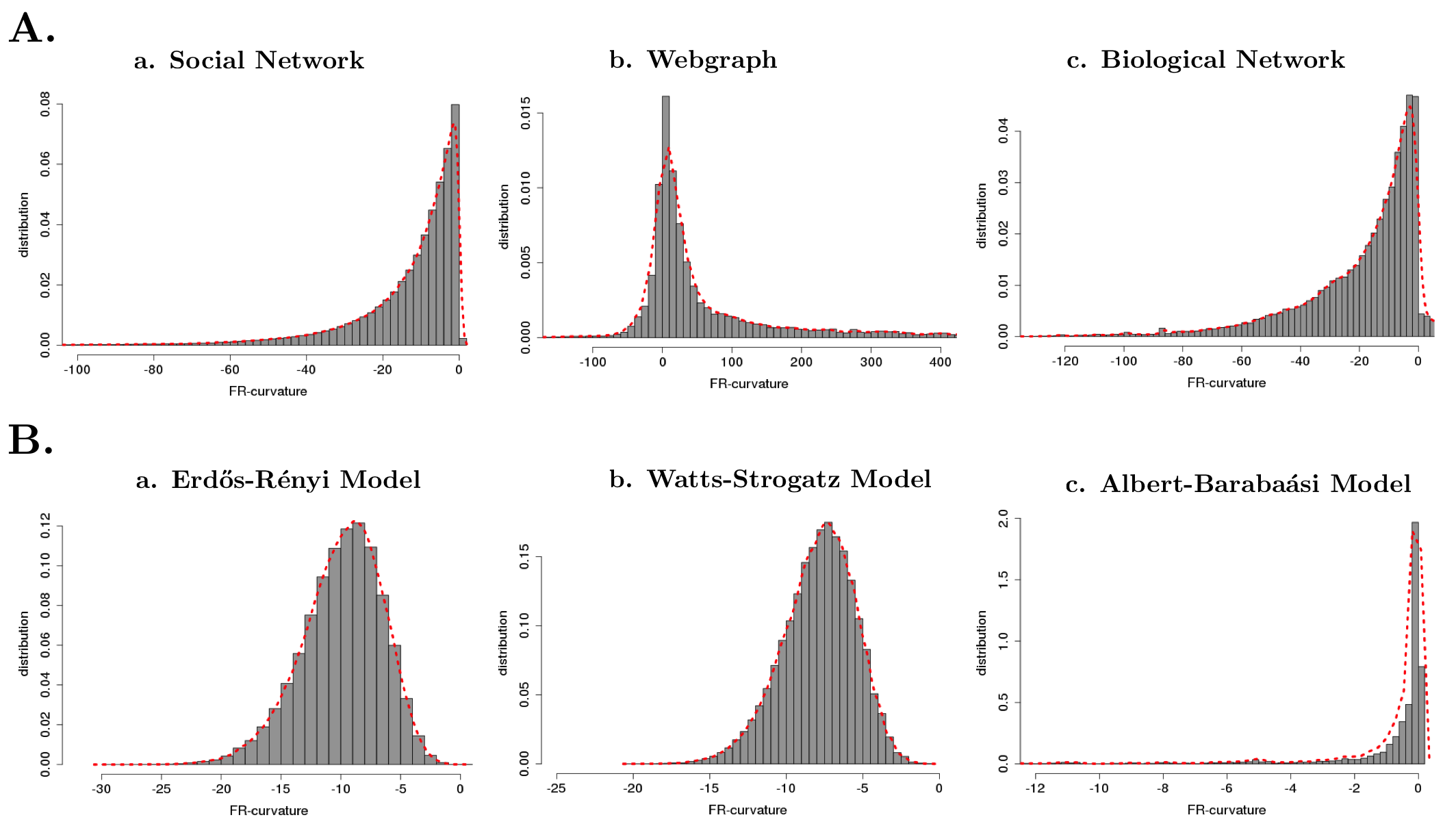} 
	\caption[Distribution of Forman-Ricci curvature with density estimates]{Results for Forman-Ricci curvature: Distributions and density curves. The three real-world examples show a close resemblance with the Albert-Barabási model with the typical power-law distribution. Notably, the width of the distribution is larger than in the case of the AB-model, hinting that real-world-networks are 'mixed types' of model networks, as previousely discussed in the literature \cite{marcotte}.}
	\label{fig_FR_h}
\end{figure}

\begin{figure}[t]
	\centering
	%\captionsetup{width=.9\linewidth}
	\includegraphics[width=\linewidth,natwidth=1652 ,natheight=1075]{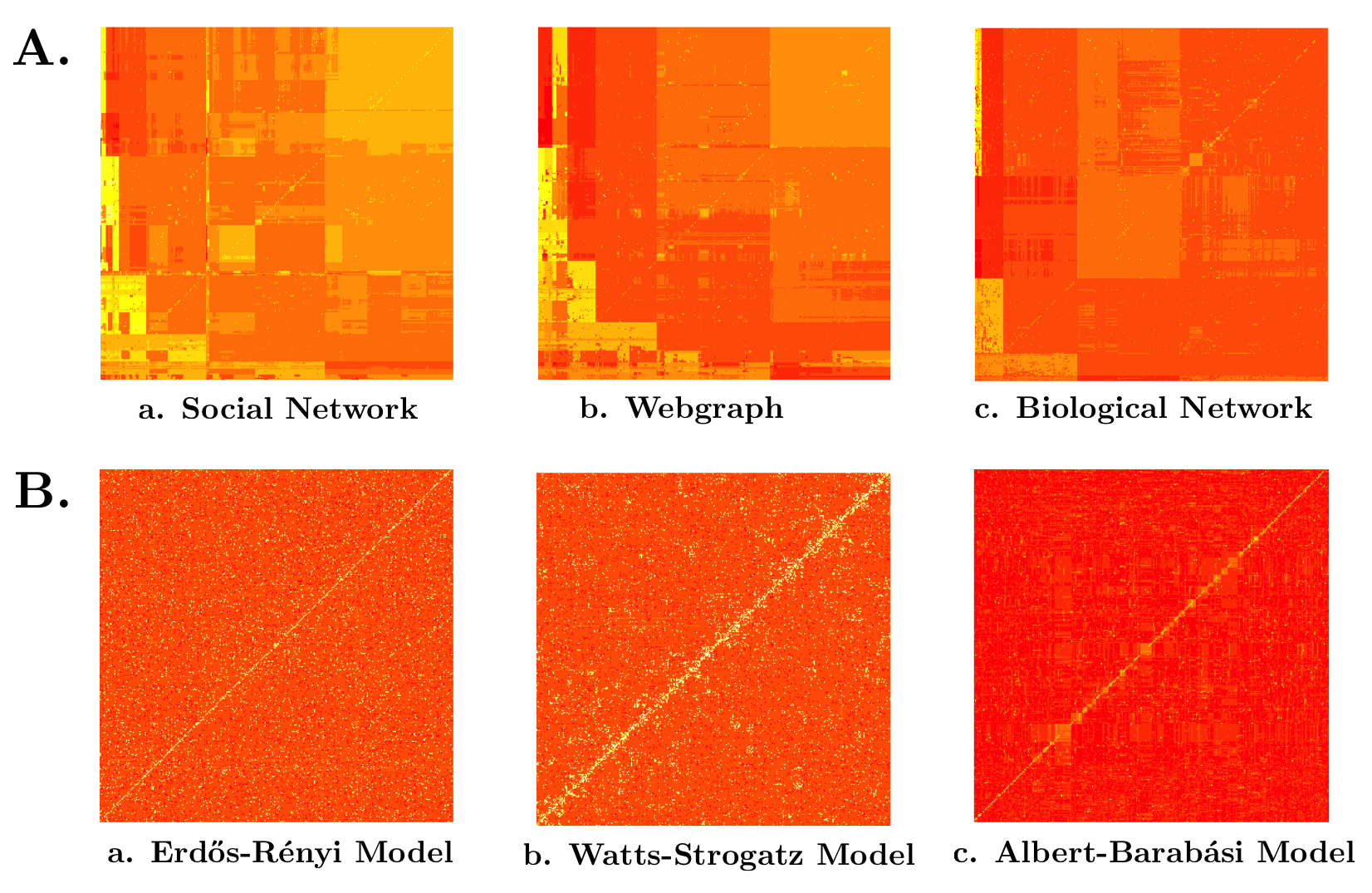} 
	\caption[Curvature maps]{Curvature maps for Forman-Ricci curvature. \textbf{A:} Curvature maps for subsamples\footnote{see \textit{Supplemental Material}} of the three real-world examples with 1,000 nodes. Both the community structure and the directionality of the networks is clearly represented in the maps. \textbf{B:} Curvature maps for the three model networks. The Albert-Barabási model (c) shows resemblance with the real-world networks due to its community structure, a result of its intrinsic scale-freeness.}
	\label{fig_FR_m}
\end{figure}

We computed\footnote{see \textit{Supplemental Material} for details on implementation and the publicly available code} the distribution of the Forman-Ricci curvature for the three real-world examples described above, as well as computationally generated examples for three model networks ($\sim 20,000$ nodes each, see \textit{Supplemental Material} for details). The results are displayed as histograms and estimated density functions (Fig. \ref{fig_FR_h}) and as curvature maps (Fig.\ref{fig_FR_m}), respectively. To quantify the dissimilarity, between the networks, we use the curvature-based distance measure introduced in the previous section. For computational purposes, the density estimates and separation into bins can be chosen in direct analogy to the histograms.
%For reference, classic characteristics of the six networks are shown in table Tab. (\ref{tab:classic-char}).
%

\noindent A comparison of the distributions (Fig.\ref{fig_FR_h}) indicates strong qualitative similarity between the Albert-Barabási (AB) model and the real-world networks. This hints on a correlation of the curvature distribution and \textit{scale-freeness}, the central property in the AB-model. A more extensive statistical analysis in \cite{SMJSS,SJSS} obtained comparable results, empirically showing a strong correlation between scale-freeness and the distribution of Forman-curvature over a wide range of network examples. 
%\add[MW]{Remove? But then, we have not much new in this section:} 
%\add[ES]{No, I think it's OK to keep this.}

To quantify this observation, we compute the distance between the estimated curvature distributions of our samples and a simulated model network of comparable size. The computation is based on the curvature-based distance introduced in the previous section and utilizes the earth mover's distance determined by the Hungarian algorithm (see \textit{Supplemental Material}). Table \ref{tab:dist} summarizes the results. As expected from the qualitative evaluation of the histograms, the Albert-Barabási model shows by far the most resemblance with the three real-world networks. Surprisingly, our results suggest, that the random graph model (Erdős-Rényi) gives a closer approximation than the Watts-Strogatz model - in contrast to what one would expect from previous studies on other network-characteristics \cite{newman1}. A closer examination of curvature-distances between a larger number of real-world networks is beyond the scope of this article and remains for future work.
\begin{table}[h]
\begin{center}   
\begin{tabular}{ p{6cm} p{6cm} }\hline
   % \hline
    \textbf{Network} & \textbf{Curvature-based distance} \\ \hline
   Facebook (social) & $d_{AB}=0.0495$ \newline $d_{WS}=0.3315$ \newline $d_{ER} = 0.2706$ \\ %\hline
   & \\
   Google (web) & $d_{AB}=0.0546$ \newline $d_{WS}=0.3581$ \newline $d_{ER} = 0.2910$ \\ %\hline
   & \\
   BioGrid (biological) & $d_{AB}=0.0604$ \newline $d_{WS}=0.2238$ \newline $d_{ER} = 0.1688$ \\ %\hline
   & \\
   Distances between \newline real-world networks & $d(F,G)=0.0565$ \newline $d(F,B)=0.0086$ \newline $d(G,B)=0.1131$ \\ \hline
    \end{tabular}
 %   \captionsetup{width=.9\linewidth}
    \caption[Curvature-based distances]{Curvature-based distances between analyzed data sets. The results show, in accordance with our qualitative observations, a close resemblance between the real-world networks and the Albert-Barabási model. The distances among the real-world networks are approximately of the same order.}
    \label{tab:dist}
    \end{center}
\end{table}

%
%\begin{itemize}
%\item discuss curvature vs. scale-freeness
%\item compute distances of probability distributions?
%\end{itemize}
%
An explanation for the observation of the close resemblance with the Albert-Barabási model follows from the geometric meaning of the curvature: Highly curved edges accumulate in fast evolving network regions that are distinguished by their information content (in the classic Riemannian analogy, large deviations from local flatness of the Euclidean sphere). Previous work in Network Analysis has shown, that major properties of networks are governed by a few densely connected, distinguished nodes (so-called hubs) and the communities of nodes that center around them \cite{newman2}. This structure is intrinsic to the scale-freeness of the Albert-Barabási-model. Our results suggest, that these hubs and densely connected communities surrounding them, are characterized by high curvature - offering an explanation for the high correlation of curvature distribution and scale-freeness. 

Additional insight into the community structure of the networks is gained from evaluating the curvature maps (Fig.\ref{fig_FR_m}A). The curvature map representation highlights the similiarity of curvature in a specific network region. The rich structure visible in the maps suggests curvature-based clustering as a possible application of the Forman curvature (and the curvature map representation), as we will discuss later in the article.

%Similar results are obtained from evaluating the curvature maps (Fig.\ref{fig_FR_m}). The maps display a community structure that is characteristic for real-world networks \cite{newman2} and also captured in the AB-model. The rich community structure visible in the maps suggests a possible application of the Forman curvature (and the map representation) for clustering. 

\subsection*{\textbf{Curvature maps and Directionality} }

\noindent Furthermore, the curvature maps offer insights into the \textit{directionality} of networks: Undirected networks have symmetric curvature maps; whereas directed ones (as the examples in Fig.\ref{fig_FR_m}A) are asymmetric. Directionality is an intrinsically edge-based network property that cannot be captured with the standard characteristics. Looking at different classes of real-world networks, including the exemplary ones studied in this article, the great importance of the directionality becomes clear:

Social interactions as well as information flow are rarely symmetric - people in leading positions tend to receive far more emails and status updates from their group members than they send themselves. Hyperlinks between web pages are highly asymmetric as, for instance encyclopedias like Wikipedia link a lot of pages through citations and suggestions for further reading - usually without being linked back by the respective page. Even if links in both directions exist, they need not be used with the same frequency, possibly resulting in different weights. Another example for which directionality is especially well studied, is that of neuronal networks. In characterizing the behavior of neurons as excitatory or inhibitory, opposite signs are assigned to the edge weights - introducing directionality to the network. Furthermore, it is well known that the information flow between different brain regions is highly asymmetric, mostly occurring in only one direction. 

Providing a tool to capture and study this intrinsic property of real-world networks is one of the major innovations of the Forman-Ricci curvature. The additional spatial dimension of the curvature maps allows for an elegant representation of directionality and its analysis in terms of community structures.

\begin{figure}[t]
	\centering
%	 \captionsetup{width=0.9\linewidth}
	 \includegraphics[width=\linewidth,natwidth=1599,natheight= 1614]{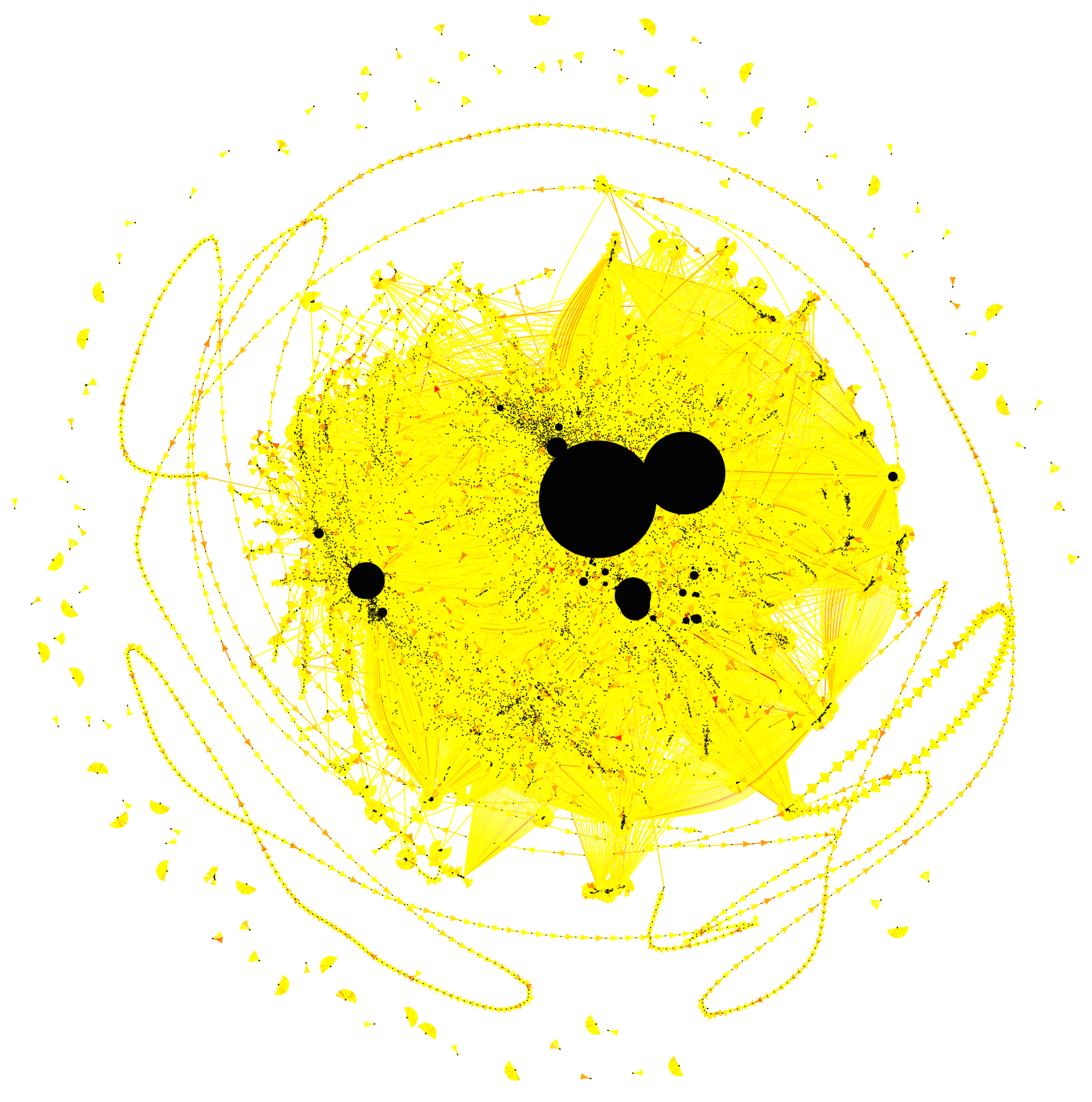} 
	\caption[Curvature-colored plot of a Google wabgraph]{Curvature-colored network plot of a Google webgraph \cite{konect}\cite{google}.}
	\label{fig:google}
\end{figure}
%
%%%%%%%%%%%%%%
%%% Ricci-Flow %%%
%%%%%%%%%%%%%%

\section{Ricci Flow on networks}

The Forman-curvature resides at the base of the very definition of the Forman-Ricci flow and is furthermore closely related to the Laplacian flow. Both flows can be utilized to characterize the dynamics in evolving networks, as we shall see in this section. In this setting, network regions (or communities) with high curvature (in Fig.\ref{fig_FR_m}: red) are fast evolving, with information flow \textit{towards} (growing with fast accumulation of nodes, positive curvature) or \textit{away from} (shrinking or fast expansion, negative curvature) them. On the contrary, in regions with low curvature (in Fig.\ref{fig_FR_m}: yellow) the flow is overall low leaving the region static or only slowly evolving.

\subsection{Forman-Ricci flow}
%\add[MW]{Definition and motivation}

\noindent Pioneered by Hamilton \cite{Ha}, and fully developed by Perelman \cite{Per1, Per2}, the Ricci flow has become by now an established and active subject of study in Differential Geometry -- see e.g. \cite{CK, To} and the references therein. 

In the special case of surfaces, the Ricci flow has a very simple and intuitive form, due to an essential identity relating Ricci curvature and the classical Gaussian curvature $K$, namely:
\begin{equation}  \label{eq:RicciFlow}
{\frac{\partial g_{ij}}{\partial t} = - K (g_{ij}) \cdot g_{ij}}\,;
\end{equation}
%
%\add[ES]{In Geometry the notation is $K$, not $\mathcal{K}$.}

\noindent This form of the flow was  adapted by Chow and Luo \cite{CL} in their work on the combinatorial Ricci flow on $PL$ surfaces. In turn, it is this equation given in term of the lengths of the edges (1-skeleton) of the given surface that inspired our own definition for dynamically evolving weighted networks in \cite{WJS}:
\begin{equation} \label{eq:RicciFlowNtwks-ch1}
\tilde{\gamma} (e)  - \gamma (e) = - {\rm Ric}_F (\gamma (e)) \cdot \gamma (e)\,;%
\end{equation}
where $\tilde{\gamma} (e)$ denotes the new (updated) value of $\gamma (e)$ after one time step.  Note that in our discrete setting, lengths are replaced by the (positive) 
%\add[ES]{Did we (me) specify the weights are positive?!...}\add[MW]{yes, see previous section} 
edge weights. Time is assumed to evolve in discrete steps and each ``clock" has a length of 1. 

In contrast, the Forman-Ricci flow based on a discretization of the ``proper'' 3- (and higher) dimensional flow, i.e.
	\begin{equation}  \label{eq:RicciFlow1}
	{\frac{\partial g_{ij}}{\partial t} = - {\rm Ric} (g_{ij})}\,;
	\end{equation}
would give us the following alternative for (\ref{eq:RicciFlowNtwks-ch1}):
	\begin{equation} \label{eq:RicciFlowNtwks1}
	\tilde{\gamma} (e)  - \gamma (e) = - {\rm Ric}_F (\gamma (e))\,.%
	\end{equation}

\begin{rem}[\textit{Choice of the form of the Ricci flow}]
As discussed in \cite{WJS}, we consider the Ricci flow in the specific form (\ref{eq:RicciFlowNtwks-ch1}) the Ricci flow has in dimension 2. However, our choice is, by no means, unique. Besides easy computability, a main reason for choosing Eq. \ref{eq:RicciFlowNtwks-ch1} is the fact that the most successfull discrete Ricci flow, namely that of Chow and Luo \cite{CL}, is of this type. 

%A main reason for our choice is both the fact that the  most succesfull discrete Ricci flow, namely that of Chow and Luo \cite{CL} is of this type. 
%
%We follow Chow and Luo \cite{CL} to choose the specific form (\ref{eq:RicciFlow}) of the Ricci flow in dimension 2. As we pointed out in \cite{WJS}, it is based on the specific form (\ref{eq:RicciFlow})  the Ricci flow has in dimension 2. One main reason for our choice is both the fact that the  most succesfull discrete Ricci flow, namely that of Chow and Luo \cite{CL} is of this type. 

While defined for triangulated surfaces (i.e. of images), it effectively resides solely on the 1-skeleton, thus providing a formalism that can be easily mapped to the network graphs we are concerned with in the present article. Moreover, this approach is also adopted in defining the only other established Ricci-type flow for networks, namely the \textit{Ollivier-Ricci flow}. It is therefore a natural choice to define a Forman-Ricci flow in a 2-dimensional form, allowing for consistency between and comparison of the two flows. However, note that considering the curvature at the nodes, the Ollivier-Ricci flow represents a {\em scalar} flow, rather than a  Ricci flow \cite{Allen1}. In contrast, our proposed form for the Forman-Ricci flow represents a true Ricci flow, since it describes an evolution prescribed by the Ricci curvature of the edges. Thus, it is in concordance with the classical (smooth) Ricci flow. 

Another, more theoretical reason for our choice resides in the fact that graphs (and hence networks), while formally 1-dimensional objects, are ``almost 2-dimensional''. In a sense, they encode in many important cases some essential properties of surfaces \cite{HJL} and, as such, hold in common with 2-dimensional structures many important features. In fact, the classic 1-dimensional graphs can be extended to higher dimensional structures intuitively, as we will discuss in a follow-up article.

However, beyond this theoretical motivation, practical reason also implies a preference for the form (\ref{eq:RicciFlowNtwks-ch1}): As experiments with images show \cite{SSAZ}, (\ref{eq:RicciFlowNtwks1}) it is in computational analysis more unstable and tends to produce certain singularities. Given the more discrete nature of networks compared to images, we must expect the flow to be even more prone to such a lack of stability. Therefore, to avoid sharp changes in curvature and weights, it is useful to benefit from the smoothing effect of the edge weight factor in the right side of (\ref{eq:RicciFlowNtwks-ch1}).
\end{rem}
%
%	However, beyond this theoretical motivation of our prefeerence that we discussed above, there exits also a practical reason for preffing the form in (\ref{eq:RicciFlowNtwks}): As experiments with images show -- see  \cite{SSAZ} and the references therein -- this version of the flow is more unstable and tends to produce certain singulatities. Given the more discrete nature of the data in our case, the flow seemingly would be even more prone to such lack of stability. Therefore, to avoid sharp changes in curvature and weights, it is useful to benefit from the molifiyng effect of the edge weight factor in the right side of (\ref{eq:RicciFlowNtwks}).
%	)
%\end{rem}
%
%\add[ES]{Is it enough/good to make this as a remark? Also, I think this is the most natural place to insert this, but I'm not absolutely sure.}
%
\begin{figure}[H]
	\centering
%	 \captionsetup{width=0.9\linewidth}
	 \includegraphics[width=0.8\linewidth,natwidth=1744,natheight= 642]{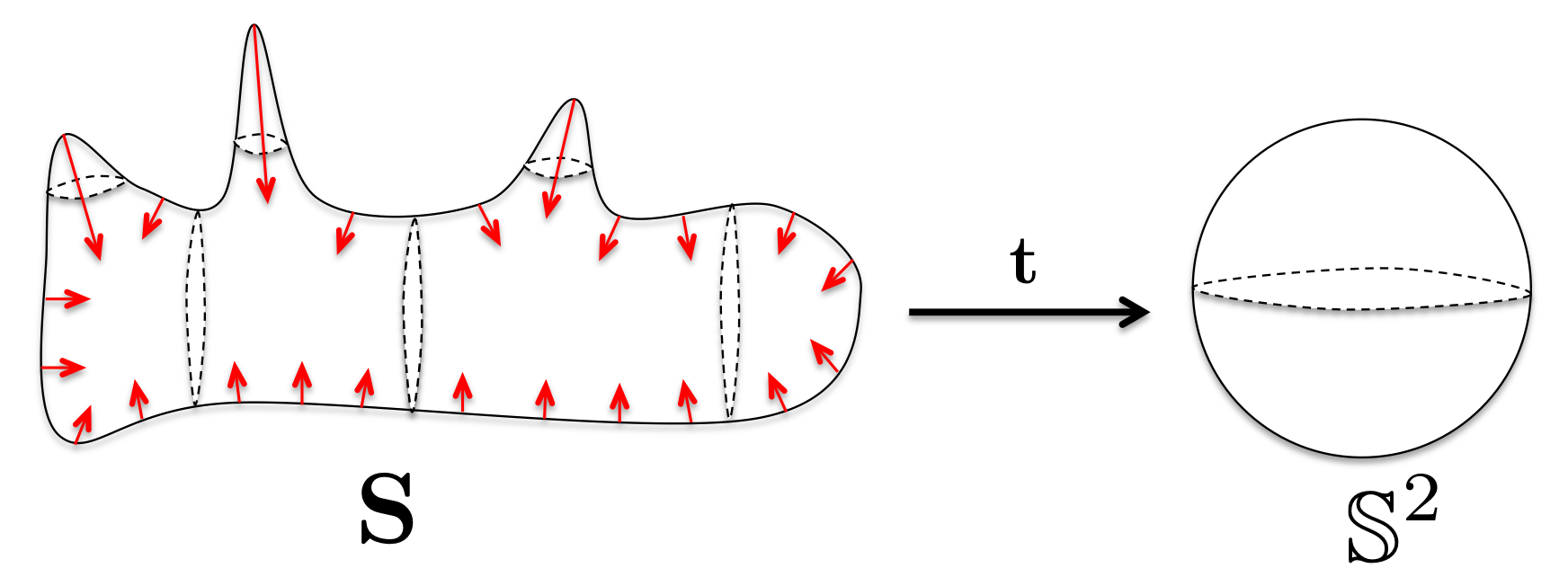} 
	\caption[Ricci flow on surfaces]{The surface Ricci flow evolves the surface proportionally to the Ricci curvature, towards a gauge surface of constant Gaussian curvature. The spherical form evolves by ``pushing'' the points of high curvature faster then those of low curvature, such that regions of positive curvature (tend to) shrink, while those of negative curvature (tend to) expand.}
	\label{fig_flow}
\end{figure}
%
%\noindent The classical Ricci flow for surfaces -- and the fact that it represents a Laplacian flow in the metric $g$, points to the expected properties of the Ricci flow for graphs. More specifically,  the surface Ricci flow evolves the surface proportionaly to the Ricci curvature, towards a gauge surface of constant Gaussian curvature equal to the mean Gaussian curvature of the given surface, i.e. a surface of constant curvature having the same {\em Euler charcateristic} as the original one. This has two aspects: 

\noindent The classical Ricci flow for surfaces gives an intuition on the expected properties of the Ricci flow for graphs. More specifically, we expect the surface to evolve through the Ricci flow proportionally to the Ricci curvature towards a model (or ``gauge'') surface of constant curvature. This intuition is based on the classical case for 3-manifolds, where  the Gaussian curvature is in the limit equal to the mean Gaussian curvature of a given surface, i.e. a surface of constant curvature having the same {\em Euler characteristic} as the original one. This has two aspects: 
\begin{enumerate}
\item The {\it short term flow} tends to smoothen the surface by ``pushing'' the points of high curvature faster then those of low curvature. This suggests some immediate applications of this flow, namely denoising (or -- for surfaces -- smoothing). Note that Eq. (\ref{eq:RicciFlowNtwks-ch1}) is, in fact an ODE and therefore reversible. The resulting {\it reverse flow} 
\begin{eqnarray}
\frac{\partial \gamma(e)}{\partial t} = \bold{+} \rm{Ric_F} \left( \gamma(e) \right) \cdot \gamma(e) \; ;
\end{eqnarray}
can also be exploited for the very opposite goals, e.g. adding of noise (or, in the case of images, sharpening). Such goals are attainable without further theoretical work.
\item One would also like to study the {\it long term flow}. The motivation for this resides in the hope, that in concordance with the classical Ricci flow for surfaces and with its discrete counterpart introduced by Chow and Luo, the flow will evolve a given network to a {\it gauge} (or \textit{model}) \textit{network} of a defined topological type (see \cite{Ha}, \cite{CL}, respectively). However, to avoid the possibility of the network ``collapsing'' under the flow, one should consider not the standard flow, but rather its {\it normalized} version. For a surface this has the following form:
\begin{equation}  \label{eq:NormalizedRicciFlow-ch1}
{\frac{\partial g_{ij}}{\partial t} = - \left(K (g_{ij}) - \overline{K}(g_{ij})\right) \cdot g_{ij}}\,;
\end{equation}
where  $\overline{K}$ denotes the mean Gaussian curvature of the given surface $S^2$, namely 
\[\overline{K} = \frac{\int_{S^2}KdA}{\int_{S^2}dA}\,;\]
($dA$ denoting the area element of $S^2$).

%\add[ES]{Not sure if "frac" gives the best form in the notation above or "/".}

By analogy with (\ref{eq:NormalizedRicciFlow-ch1}) we can define the following {\it Normalized Forman-Ricci flow} as
\begin{eqnarray} \label{eq:NormalizedFormanRicciFlow}
\frac{\partial \gamma(e)}{\partial t} = \left( \rm{Ric_F} \left( \gamma(e) \right)  - \rm{\overline{Ric}_F}\right) \cdot \gamma(e) \; .
\end{eqnarray}
Such a \textit{``gauge" theory} would be extremely important for the study of the evolution of networks and moreover their prediction. Here, we restrict ourselves to empirical tools and thus only perform relevant experiments for the short term flow. In a follow-up article we will attempt to study such \textit{gauge networks} with theoretical tools for higher order network structures. For some additional remarks, see the discussion section of the present paper.
\end{enumerate}
%\add[ES]{I think this is more than enough - there are books about the subject, and this is not the place to expan too much on the subject.}
%
%\add[MW]{EMIL: Adress metric issue. Define (?) "pseudo" metric.}
%\begin{rem}[\textit{Scalar Ricci flow}]
%By analogy to the classic case, we can also define a {\em scalar Forman (curvature) flow} (see \citep{WJS}, Eq. (9)). 
%%(One should note again in this context that, in previous works on Ollivier-Ricci curvature for networks \cite{Allen1}, \cite{NLGGS}, this scalar curvature was actually considered, rather than the professed Ricci one.) 
%\end{rem}
%%
\begin{rem}[\textit{Metric structure}]
We note that, while in its present form formally correct and applicable to the cases considered here, an extension to the long term flow is a relevant open point to address in future work. An essential aspect in this context is the introduction of a metric structure: For more rigorously founded geometrical conclusions, such as the evolution of the network under the long term flow, one needs to ensure that the edge weights endow the network with a metric structure. Indeed, without further assumptions, the given weights represent a distance function only on the {\it star} of a node, i.e. on the edges adjacent to it (and their end nodes). One approach for remediating this issue is the consideration of the {\it path metric} induced by the given weights. 
%\add[MW]{Emil: please add a reference for the path metric of you have}
	%\add[ES]{Should I add a reference for this metric? - I did, but perhps you won't find this necessary.}
\end{rem}

\subsection{Laplacian Flow}
%\add[ES]{"subsubsection" only or it deserves a full "subsection"?} 
%\add[MW]{This should be a separate sub-section that also features results from experimental comparison.}
%\subsubsection*{Definition for networks}

Through the \textit{Bochner-Weitzenb\"{o}ck formula}, Forman's Ricci flow is closely related to the \textit{Laplace-Beltrami} (or short \textit{Laplacian}) \textit{flow}. In this section, we want to give a network-theoretic formulation of the Laplacian flow and its relation to the Forman-Ricci flow. 

We start with the general notion for Riemannian manifolds and develop a formulation for graphs through an analogy for grids and manifolds that has been successfully applied in Imaging (see, e.g., \cite{Xu,KMM}).

When considering more general types of surfaces, the general Laplace-Beltrami operator on a Riemannian manifold $(M^n,g)$ is given by (\cite{Be}):
\begin{equation} \label{eq:Laplacian-Manifolds}
	\triangle_g  = \; \frac{1}{\sqrt{{\rm det}(g_{ij})}}\sum_{i,j = 1}^n\frac{\partial}{\partial x^k}\left(\sqrt{{\rm det}(g_{ij})}g^{i,j}\frac{\partial I}{\partial x^l}\right) \;,
\end{equation}
where $g = (g_{ij})$ and $g^{ij} = g(dx^i,dx^j)$.
\begin{rem}
Note, that in contrast to the prevalent convention in Riemannian Geometry, but according to the intuition (and common practice in the Applied Sciences), we have opted for the sign ``+'' in the definition of the Laplacian, as well as in that of the Ricci flow. 
\end{rem}

In analogy to a manifold \cite{Xu,KMM}, we consider an image $I$, viewed as a real, positive function defined on a rectangle (the image base). Then the Laplace-Beltrami flow is defined by
\begin{equation} \label{eq:LaplacianFlow-Images}
	\frac{\partial I}{\partial t} \; = \; \triangle_g I \;.
\end{equation}
where $\triangle_g I$ is the standard Laplacian of the metric $g$ of $I$. The metric $g$ itself is  determined by the 1-skeleton of the image, i.e. by the edges between the $I$-images of adjacent pixels. (For more details see, for instance, \cite{ASZ} and the references therein.) 

Since the edge grid of an image is nothing but a very simple, regular graph, except at the boundary, it is quite natural to extend this type of flow to weighted graphs and networks. We substitute $\triangle_g$ with the Bochner Laplacian obtaining a simplified version of the Laplacian flow for graphs:
\begin{equation} \label{eq:RoughLaplacianFlow}
\frac{\partial G}{\partial t}  =  \Delta_F^1 G\,.
\end{equation}
%
%where $E := E(G)$ denotes the set of edges. 
From (\ref{eq:RoughLaplacianFlow}) and from the Forman version of the Bochner-Weitzenb\"{o}ck formula (\ref{saucan-eqn:3}), we easily obtain a fitting version for the Laplacian flow on networks:
\begin{equation} \label{eq:BochnerLaplacianFlow}
\frac{\partial G}{\partial t}  = \Delta_F^1 G = (\Box_1 - {\bf F})G\,.
\end{equation}

\noindent For the notation in the defining expression (\ref{eq:RoughLaplacianFlow}), we follow the classic convention in Imaging (\ref{eq:LaplacianFlow-Images}). However, note that in practice the flow ``resides'' on the edges of $G$, and the evolution of the edge weights therefore intrinsically determines the eventual evolution of the nodes. 
Note that in both cases the absolute value of the edge weights $\gamma$ %\add[ES]{Consistent notation?...} 
generalize the lengths of the edges (and the metric $(g_{ij})$). Given (\ref{saucan-eqn:Lap}) and the Bochner Laplacian 
%
%\add[ES]{Discussion here about how to produce a true metric or better not to complicate ourselves?...}
%
%\add[MW]{Please add.}
%
%\add[ES]{I just did at the end of Subsection 4.1 - move it wherever you thik it fits the best.}
%\add[ES]{All this as a Remark or in text, as it is now?}
%
\begin{align}
\Box_1 (e) = \sum_{e \sim v} \frac{\omega (v)}{\gamma (e)} \; ;
\end{align}

\noindent from (\ref{weber:FormanLap}), we can calculate the Laplacian flow, according to (\ref{eq:BochnerLaplacianFlow}).

%the two flows obtain, in our case the following forms: 
%
%\begin{equation}
%...
%\end{equation}
%\add[ES]{Insert here the formula according to the expression of $\Box_1$ that you actually used.}
%
% respectively
%
%\begin{equation}
%...
%\end{equation}
%\add[ES]{ICombine the formula according to the expression of $\Box_1$ that you actually used with the one for $F$.}

\subsection{Characterizing dynamic effects in networks}
%\begin{footnotesize}
%\begin{itemize}
%\item review method/ technical details (from \cite{WJS}) \add[MW]{(y)}
%\item experimental results from gnutella-analysis
%\item comparison with Bochner-Laplacian
%\item discussion of application: extrapolation of future behavior, detection of unusual changes/ interesting regions. 
%\end{itemize}
%\end{footnotesize}

\subsection*{\textbf{Detection of changes and ``interesting" regions with Forman-Ricci flow} }

\noindent In \cite{WJS}, we introduced a formalism for characterizing dynamics in complex networks. We will review the mathematical details here and then perform a more sophisticated dynamic analysis of the Gnutella network \cite{snap}, \cite{gnutella1}, \cite{gnutella2}, going well beyond the investigations in \cite{WJS}.

The Ricci flow based on Forman curvature provides a tool for studying dynamic effects, resulting from information flow, in complex networks. For this, we consider a set of 'snapshots' that specify the network's structure (connections between nodes and their strength) at given (discrete) time points $\lbrace t_i \rbrace$ represented as weighted graphs $(G_i)_{i \in I}$. By analyzing the flow of the curvature between the time points, we gain insights into structural changes that can be utilized in a number of applications, as discussed below.

For computing the flow, we consider the following formalism: Let $t_i$ and $t_{i+1}$ be adjacent time points with corresponding graphs $G_i$ and $G_{i+1}$. The Forman-curvature characterizes the geometric structure of $G_i$ and $G_{i+1}$ based on the weighting schemes
\begin{eqnarray}
\omega_{n,n+1}^0: V(G) \mapsto [0,1] \; ,
\end{eqnarray}
and edges
\begin{eqnarray}
\gamma_{n,n+1}^0: E(G) \mapsto [0,1]  \; ,
\end{eqnarray}

\noindent at times $t_i$ and $t_{i+1}$. From this, we compute the Ricci flow Eq. (\ref{eq:RicciFlowNtwks-ch1}) at $\Delta t=t_{i+1} - t_i$ by iterating $k=1, ..., K$ times:
\begin{small}
\begin{eqnarray}
\gamma_{i,i+1}^{k+1} (e) = \gamma_{i,i+1}^{k} (e) - \Delta t \cdot {\rm Ric}_F (\gamma_{i,i+1}^{k} (e) ) \cdot \gamma_{i,i+1}^{k} (e) \; .
\label{eq:Ricci-flow}
\end{eqnarray}
\end{small}
resulting in updated weighting schemes $\gamma_i^K$ and $\gamma_{i+1}^{K}$. The correlation between the updated weights is a measure of the flow Eq. (\ref{eq:Ricci-flow}), identifying network regions that have undergone significant structural changes at $\Delta t$ by evaluating the corresponding (thresholded) correlation matrix. Alternatively, one can align the edges of $G_i$ and $G_{i+1}$ beforehand and compare the distance between the edge weights (using the $L^1$-norm). We will use the later one here, for the correlation-approach see \cite{WJS}.

For computing\footnote{see \textit{Supplemental Material} for details on implementation and the publicly available code} the Ricci flow, we use the following work flow: 
%
%\noindent Following method for detecting changes in images \cite{SSAZ} using curvature, detect changes in FR-map for large data sets:
\begin{enumerate}
\item Calculate the Forman-Ricci curvature for the network's states at times $t$ and $t+\Delta t$ (e.g. version of a data base, downloaded at $t_i$ and its preceeding version from $t_{i+1} = t_i + \Delta t$ ).
\item Normalize the networks, i.e. the updated weighting schemes $\gamma_i^k$ and $\gamma_{i+1}^{k}$. 
\item Calculate the discrete Ricci flow for each network through iteration over 
\[\gamma_{ij}^{k+1} = \gamma_{ij}^{k} - \Delta t \cdot Ric(\gamma_{ij}^{k}) \; .\]
%with Forman-Ricci curvature $Ric_F$ ($\gamma^0$ refers to the original weighting scheme of the network).
\end{enumerate}
%\add[MW]{'old' comment of yours in the draft of the conference paper ... should we adress this here?}
%\add[ES]{ I think that here we should either explain that this is a short-time flow (but perhaps not say precisely how we compute...) or use the correct formula - and apply it (but for this is no time...) I'll add this myself once we decide.} \\
%
% figure panel
\begin{figure}[htb]
	\centering
%	\captionsetup{width=0.9\linewidth}
	\includegraphics[width=\linewidth,natwidth=2118 ,natheight= 1645]{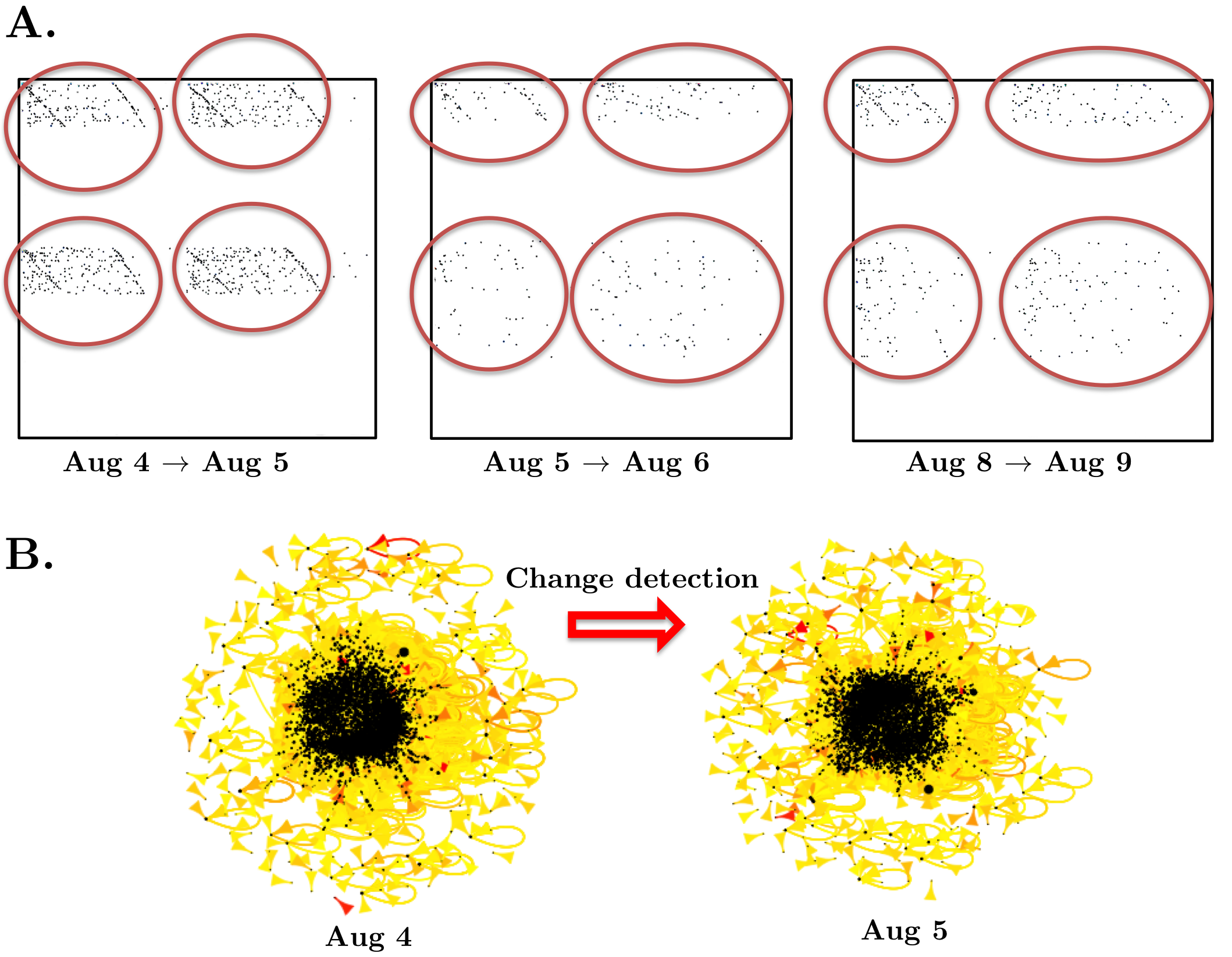} 
	\caption[Change detection in the Gnutella-network]{Characterization of information exchange in Gnutella. \textbf{A:} Change detection with Forman-Ricci flow and $L^1$-distance (K=10, threshold=0.1). \textbf{B:} Graph plot of Gnutella information (August 4 and August 5 2012). Edges are colored according to their Forman-Ricci curvature.}
	\label{fig:gnutella}
\end{figure}

\noindent Fig. \ref{fig:gnutella} shows the detection of fast evolving edges and, more broadly, the identification of ``interesting" dynamically changing regions. The capability of the Ricci flow to detect such regions relates to our observations on the Forman-Ricci curvature in static networks in the previous section: Hubs in dense network regions (communities) govern the global properties of the network through strong connections with their neighbors. Those strong connections are characterized by high (absolute) Forman curvature, which is by its very definition, related to network growth. Regions of high curvature are therefore fast evolving, the respective changes in edge weights -- characterized by the Ricci flow -- describe the growth or shrinkage behavior.

%\add[MW]{Results + Evaluation gnutella}

\subsection*{\textbf{Denoising with Laplacian flow}}
%\add[MW]{Proposal for denoising, useful for second paper also. Basic challenge: How to reconstruct networks with the Laplacian flow?}

\noindent Networks built from empirical data, such as the ones considered in the previous section, are greatly affected by noise in the underlying data. Especially sensitive are the edge weights that describe the strength of the inferred associations. Utilizing the \textit{averaging effect} of the Laplacian flow, we propose a method for correction of arbitrary small fluctuations as a first step towards denoising networks. When inferring a network from given data, and assuming, that small fluctuations mainly affect the strength of the edges and not their existence itself, one gets
\begin{eqnarray}
\tilde{G} = \lbrace V, E, \omega, \tilde{\gamma}	\rbrace \; ,
\end{eqnarray}

\noindent Under this assumption, we can limit our analysis to the denoising of the edge weights, i.e. construct
$\gamma$ from $\tilde{\gamma}$:
\begin{eqnarray}
\tilde{\gamma} = \gamma + dt \cdot dW \; .
\end{eqnarray}

\noindent We assume arbitrary small fluctuations of the form of \textit{Gaussian white noise} $dW$ and attempt to smooth the resulting noise effects with the Laplacian flow:
\begin{eqnarray}
\gamma \approx \tilde{\gamma} - dt \cdot \Delta_F^1 (G) \; .
\end{eqnarray}

\noindent This formalism is restricted to cases where the correction term is small with respect to the edge weights. Additionally, we only consider the short term flow, assuming that the fluctuations are mainly a result of spontaneous irregularities that occur over a short term.

\begin{figure}[htb]
	\centering
	% \captionsetup{width=0.9\linewidth}
	 \includegraphics[width=\linewidth,natwidth=1952 ,natheight= 1402]{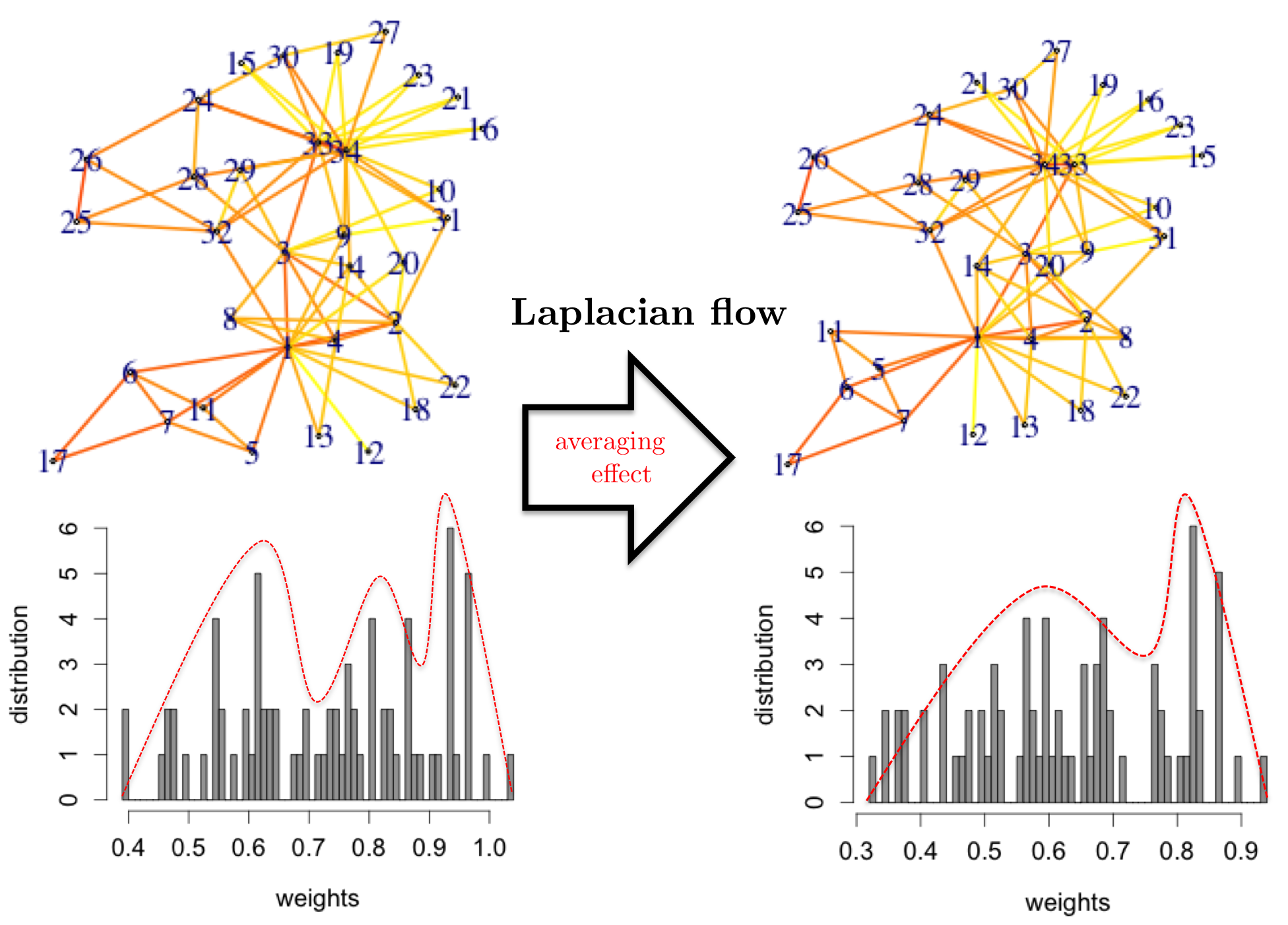} 
	\caption[Denoising with Laplacian flow]{Denoising with Laplacian flow. The \textit{averaging effect} of the Laplacian flow smoothes the distribution of edge weights and corrects for small fluctuation and random noise in the underlying data (here: Zachary's karate club \cite{karate}). For illustration, we highlight the averaging effect schematically.}
	\label{fig:lap_flow}
\end{figure}

With the corrected weights $\gamma$, we can reconstruct the denoised graph as
\begin{eqnarray}
G= \lbrace V, E, \omega, \gamma \rbrace \; .
\end{eqnarray}
%
%
%\subsubsection*{Use case}
%
\noindent Fig.\ref{fig:lap_flow} shows the correction of fluctuations in the edge weights with the Laplacian flow for a small use case. The intuitive scheme of the method can be easily extrapolated to large networks. The distance between the original and the corrected weights (i.e. the \textit{denoising level}) can be calculated using the $L^1$-norm, since the edges are aligned. First (preliminary) computational tests on a selected set of real-world networks showed promising results. %\add[MW]{presentation of results?}

\subsection{Duality of Ricci curvature and the Laplacian}
%\add[MW]{We discussed this point in a meeting before you left, it might be an interesting point to adress in this context.}

%\noindent The derivation of formula (\ref{saucan-eqn:1}), hence of the definition of the Forman-Ricci curvature (\ref{eq:Forman}), from the Bochner-Weitzenb\"{o}ck formula (\ref{saucan-eqn:3}), does not represent a mere ingenious mathematical artifice -- it has real and deep meaning, that stems from that of the classical formula,  connecting the curvature of the manifold with the Riemann-Laplace operator, hence to the heat diffusion. 

\noindent The derivation of the definition of the Forman-Ricci curvature (\ref{eq:Forman-ch1}) from the Bochner-Weitzenb\"{o}ck formula (\ref{saucan-eqn:3}) introduces a deep theoretical insight: It connects the curvature of the manifold to the Riemann-Laplace operator, hence to the heat diffusion. Loosely formulated, the Laplacian flow incorporates both an ``approximate'', or rough Laplacian, as well as a correcting geometric component, namely the curvature term. In consequence, the Laplacian flow captures the behavior of both the Laplacian on and the curvature of the underlying manifold. Note that only for the case $p = 1$, i.e. for the Ricci curvature, the curvature term has a clear and specific geometric content.

Moreover, the importance of the observation above is not restricted to the theoretical realm, but has concrete and meaningful practical consequences. Indeed, the fact that one can obtain, through Forman's method \cite{Fo}, both a Ricci curvature and a Laplacian, endows us with the discretization of two powerful geometric, respective analytic, tools, that have found a wide range of applications in various fields of Computer Science. In particular, one can consider not just the Forman-Ricci flow introduced above, but also a fitting discretization of the Laplacian flow, which we discuss in more detail below.

%\add[MW]{Needs improvement, partly move to "Ricci flow vs. Laplacian flow".}
However, let us first emphasize the main significance of this \textit{duality} between curvature and Laplacian. Clearly, it allows us to define discrete, network-theoretic versions of both the Ricci and the Laplacian flow. This, in turn, allows for capturing a number of network properties through the specific characteristics of the two flows.

Curvature and, in particular, Ricci curvature, is an edge-based network characteristic and therefore naturally associated with the edge structure and contours of the system characterized by sharp changes in curvature. The corresponding Forman-Ricci flow is expected to identify the equivalent of such features, i.e. regions where curvature changes abruptly over a short time. Those defining features act as the ``backbone" of the network, providing a dense synopsis of its topology and structural information. We refer to this intrinsic property of the Forman-Ricci flow as \textit{backbone effect}. It was this very behavior of the short term Ricci flow that we exploited in \cite{WJS} when first introducing the change detection method. 

On the other hand, the Laplacian flow has a strong \textit{averaging} or \textit{smoothing effect} that is intuitively implied by its origins in the heat equation. It has been successfully applied for smoothing and filtering of noise in Imaging (see e.g. \cite{ASZ}, \cite{SSAZ}). In an attempt to correct for small fluctuations in networks built from noise-effected empirical data, we transferred this property to networks, with the same noise-reducing effect. 

Computational experiments using the respective other flow yielded less promising results. Especially the Laplacian flow seems, due to its strong smoothing effect, ill-suited for the detection of sharp increases in curvature as utilized in the change detection method. Furthermore, we note that in their discrete versions, the defining equations of both the Ricci and the Laplacian flows are reversible. This is of practical importance for the first introduced method of change detection, considering that a specific reverse flow (i.e. endowed with ``-'' sign) provides a tool for different applications than the original one (with the ``+'' sign). In particular, the reverse $\Box_1$ flow functions as sharpening operator, whereas the direct flow $\Box_1$ provides an efficient anomaly detection tool.

\section{Discussion}

In this article, we applied Forman's Ricci curvature and Bochner Laplacian \cite{Fo} to networks and examined their capabilities as network characteristics. We found, that additional structural information is encoded in the edge weights, which is difficult to evaluate using the established node-based characteristics. This shortcoming affects especially the directionality of edges that is of high importance in the analysis of many real-world networks. The curvature maps introduced in this paper allow for representation of directionality, possibly providing insights into the information flow and evolutionary behavior of densely interconnected subgraphs (communities).

%For this, we applied the theoretical methods to subsamples of large real-world networks and standard types and analyzed the curvature's distribution across the network. 
%We compared our results with commonly used node-based network-analytic concepts in an attempt to evaluate the additional information that can be gained from performing edge-based analysis.

A comparison of our results for three real-world networks with established model networks confirmed the observation in earlier work \cite{SMJSS,SJSS} of a close resemblance in the curvature distribution of real-world networks and the Albert-Barabási-model. We extended the statistical approach therein by introducing a curvature-based distance measure between graphs utilizing the earth mover's distance between the respective curvature density estimates. This formalism gives rise to a classification scheme for networks based on Forman-Ricci curvature as we demonstrated for a small set of samples.\\

\noindent One could argue that there are other approaches to define a curvature on networks than Forman's that have been studied more intensely, including Ollivier's \cite{Shiping,NLGGS,Allen2}. One caveat of Forman's curvature in relation to Ollivier's resides in its intrinsic local definition being defined only for edges, i.e. adjacent pairs of nodes. However, one can construct order-2, -3, etc. {\em neighbourhood graphs} by connecting all nodes that are at combinatorial distance 2 from the given node. The  weight associated to this ``order 2" edge would be the sum of the weights of the composing (original) edges (for the combinatorial distance this would equal 2). Now one can compute the Forman curvature of the resulting weighted graph. In addition, there are strong indications that Forman's and Ollivier's curvature are closely interrelated (as we shall shortly show in a forthcoming study). While measuring similar aspects of network topology, Forman's is, however, much easier and efficient to compute. This is especially important for the typically large networks relevant for applications to real-world complex systems.

In the second part of the article, we extended our approach to dynamic networks by applying the Forman-Ricci flow - based on Forman's Ricci curvature - to networks. We introduced a method to detect changes in an evolving network. Furthermore, we suggested the Laplacian flow as a tool for denoising networks built from empirical data.

While only theoretically proposed and illustrated solely on a small example in this work, a number of practical applications follow immediately from the theoretical results:
\begin{enumerate}
\item \textbf{Classification}\\
In the second section we introduced a curvature-based distance measure that quantifies the dissimilarity between a pair of graphs. 
Naturally, this distance gives rise to a classification scheme. We were able to reproduce and quantify previous statistical results in \cite{SMJSS,SJSS,WJS}, that demonstrated a strong similarity between different classes of real-world networks and the Albert-Barabási model network. A more extensive study of a larger number of data sets and comparison with various model networks remains subject to future work.
\item \textbf{Detection of ``interesting" regions} \\
The Ricci flow characterizes regions of significant dynamic changes in the curvature, hinting on structural changes in the underlying system. Looking at the flow over the course of time, regions of ongoing changes are likely to contain significant information about the system and its dynamics. Therefore, the flow provides a tool for identifying ``interesting" regions that are worth a closer analysis. We outlined the respective computational tools and demonstrated their abilities on a use case. Further study and a statistical analysis of a larger set of test samples is left for future work.
\item \textbf{Network extrapolation} \\
From knowledge of the flow over a given time range, one could extrapolate the dynamics of past and future time points. In analogy to the flow in the physical sense, one could model the flow with common equations of motion, yielding approximations of the long-term development of the network and prediction of future network states. For this, one would need to extend the assumption of short-term flows, used in the present article for the sake of simplicity, to the long term case. We only aimed to lay a theoretical fundament for curvature-based network extrapolation leaving extended studies for future work. 
To utilize the full potential of such a long-term flow, one needs to consistently define the ``gauge" networks mentioned when discussing the normalized flow. One serious impediment to this endeavor resides in the fact that the Forman-Ricci curvature does not satisfy a Gau{\ss}-Bonnet type theorem, thus -- in contrast with the surface case -- no clear topological class can be assigned to a gauge network. In a forthcoming article by the authors we will show how one can remedy this drawback and indeed define a long time Forman-Ricci flow.
%
%\add[MW]{This would represent an application of the long-term flow. Adress here.}
%
\item \textbf{Clustering} \\
The curvature maps introduced in section two strongly suggested a relation between community structure and curvature clusters. This is an argument for the application of community detection algorithms to the curvature distribution across the network in an attempt to improve existing clustering tools. A future computational method using this approach could build on earlier, related work for genetic networks \cite{dna_cluster}. For now, it remains a subject for future work.
\item \textbf{Denoising} \\
A natural application of the geometric flows associated with Forman-Ricci curvature is denoising. Empirical results in related fields, particularly Graphics and Imaging, suggest that the Laplacian flow gives good results for denoising. We transfer this idea to the network-theoretic Laplacian flow introduced in the present paper and perform a similar analysis for denoising of networks built from (noise-affected) empirical data. The promising results obtained for test cases motivate the further exploration of this idea in the future.
\end{enumerate}

\noindent We showed, that Forman's theoretical work \cite{Fo} allows for defining both the Ricci and the Laplacian flow for network in an intuitive and efficiently computable way. The applications proposed above utilize the intrinsic strengths of both flows: The \textit{backbone effect} of the Ricci flow allowing for identification of defining properties and dynamics, and the \textit{averaging effect} of the Laplacian flow capable of smoothing and correcting small irregularities in the underlying empirical data. The existence of these two complementary characteristics and their strong relation through the \textit{Bochner-Weitzenb{\"o}ck-formula} offers a rich basis for dynamic network analysis. \\

\noindent This article presents a curvature-based approach to characterize networks adding an edge-based characteristic to the established network-analytic ``toolbox". Building on and extending previous work of the author and the group, we presented the mathematical motivation of the curvature-based approach and provided examples for different types of real-world complex networks. Furthermore, we propose a number of practical applications, specifically utilizing Forman's Ricci flow that will be extended in future work.

With our work we hope to add to the common effort of developing new network-analytic tools for handling the challenges in data science come with the increasing amount of available data and the recent rapid development of the field.

\section{Acknowledgements}
The authors thank Areejit Samal for helpful discussions. ES gratefully acknowledges the support and warm hospitality of the Max Planck Institute for Mathematics in the Sciences, Leipzig. ES would also like to thank Eli Appleboim for many stimulating discussions on Forman's work, in particular in regard to its applications in Imaging. MW was supported by a scholarship of the Konrad Adenauer Foundation.
%\add[ES]{I think this is warm enough}


%\bibliographystyle{comnet}
%\bibliography{sample}
%
% once the .bbl file has been generated then place the text in your article.

% To get the unnumbered reference style the author should use [unnumbib]
%as an option in the document class.  For example: \documentclass[unnumbib]{comnet}

%\begin{thebibliography}{99}
%
%\bibitem{Rottmann:2010a}
%{\sc Rottmann-Matthes, J.} (2011a) Linear stability of traveling
%waves in nonstrictly hyperbolic PDES.\break {\em J. Dynam.
%Differential Equations}, \textbf{23}, 365--393.
%
%\bibitem{Rottmann:2011a}
%{\sc Rottmann-Matthes, J.} (2011b) Stability and freezing of
%nonlinear waves in first-order hyperbolic PDEs. Preprint
%11-016, CRC 701, Bielefeld University.
%
%\bibitem{Rottmann:2011b}
%{\sc Rottmann-Matthes, J.} (2012) Stability of
%parabolic-hyperbolic traveling waves. Preprint 12-005, CRC
%701, Bielefeld University.
%
%\bibitem{RowleyKevrekidisMarsdenLust:2003}
%{\sc Rowley, C. W., Kevrekidis, I. G., Marsden, J. E. \& Lust,
%K.} (2003) Reduction and reconstruction for self-similar
%dynamical systems. {\em Nonlinearity}, \textbf{16},
%1257--1275.
%
%\end{thebibliography}

\section*{Supplemental Material}
\noindent \textbf{Supplement 1:} Sampling of large networks. \\
\textbf{Supplement 2:} Approximation of EMD with total variation. \\
\textbf{Supplement 3:} Implementation of curvature and Laplacian analysis. \\
\textbf{Supplement 4:} Data sets.

%\add[MW]{Please find below some suggestions for references to cut (d) or rethink (?). We might need to extend this, I don't think we should have more than 50 references .. Please feel free to suggest further cuts.}
% ---------------------
%% REFERENCES
% ---------------------
\bibliographystyle{apalike}
\bibliography{manuscript_WeberSaucanJost_final}

\end{document}